\pdfoutput=1
\documentclass[11pt,a4paper]{article}
\usepackage{jheppub}

\RequirePackage[utf8]{inputenc}

\usepackage{amsfonts}
\usepackage{amssymb}
\usepackage{comment}
\usepackage{graphicx}
\usepackage{amsmath}
\usepackage{tabu}
\usepackage{dsfont}
\usepackage{float}
\usepackage{amsthm}
\usepackage{slashed}
\usepackage{setspace}
\usepackage[labelformat=simple]{subcaption}

\usepackage{pgfplots}
\usepackage{tikz}
\usepackage{tikz-cd}
\usetikzlibrary{shapes.misc,shadows,fit,decorations.pathmorphing,positioning,trees,decorations.markings,decorations.pathreplacing,calc,shapes,patterns,arrows}
\usepackage{xcolor}
\usepackage[boxsize=0.5em,aligntableaux=center]{ytableau}
\usepackage{makecell}
\usepackage{environ}
\usepackage[utf8]{inputenc}
\usepackage{amsmath, amsthm, amssymb, amsfonts}
\usepackage[font=small, labelfont=bf]{caption}
\usepackage{amsmath, amsthm, amssymb, amsfonts}
\usepackage{multirow}
\usepackage{graphicx}
\usepackage{float}
\usepackage[numbers,sort&compress]{natbib}
\usepackage{jheppub}
\usepackage{enumerate}
\usepackage{hhline}
\usepackage{amsmath}
\usepackage{amsfonts}
\usepackage{amssymb}
\usepackage{graphicx}
\usepackage{cancel}
\usepackage{makecell}
\usepackage{tabularx}
\usepackage{cellspace}
\usepackage{longtable}
\usepackage{mathtools}
\usepackage{layouts}
\usepackage{empheq}
\usepackage{color}
\usepackage{hyperref}
\usepackage[capitalize,sort]{cleveref}

\crefname{figure}{Figure}{Figures}
\crefname{table}{Table}{Tables}

\usepackage{float}
\restylefloat{table}
\restylefloat{figure}

\newcommand{\nn}{\nonumber}

\newcommand{\nc}{\newcommand}
\nc{\beq}{\begin{equation}}
\nc{\eeq}{\end{equation}}
\nc{\be}{\begin{equation}}
\nc{\ee}{\end{equation}}
\nc{\bea}{\begin{eqnarray}}
\nc{\eea}{\end{eqnarray}}
\nc{\bi}{\begin{itemize}}
\nc{\ei}{\end{itemize}}
\nc{\ben}{\begin{enumerate}}
\nc{\een}{\end{enumerate}}

\def\ov{\overline}

\allowdisplaybreaks[2]
\numberwithin{equation}{section}

\def\vo{\mathcal{V}}

\allowdisplaybreaks[2]
\numberwithin{equation}{section}

\title{Stabilising all K\"ahler moduli in perturbative LVS}
\author[a,b]{George K. Leontaris,}
\author[c]{Pramod Shukla}
\affiliation[a]{\small Theoretical Physics Department, CERN, 1211 Geneva 23, Switzerland}
\affiliation[b]{\small Physics Department, University of Ioannina, 45110, Ioannina, Greece}
\affiliation[c]{\small ICTP, Strada Costiera 11, Trieste 34151, Italy}
\emailAdd{leonta@uoi.gr}
\emailAdd{pramodmaths@gmail.com}

\abstract{In this work we investigate the moduli stabilisation problem and the requirements for de Sitter vacua  within the framework of type IIB string theory. Using perturbative effects arising from the various sources such as $\alpha^\prime$ corrections, logarithmic as well as KK and winding-type string-loop corrections along with the higher derivative $F^4$-contributions, we present a moduli stabilisation scheme in which the overall volume is realised at exponentially large values such that $\langle {\cal V} \rangle \simeq e^{a/g_s^2}$ in the weak coupling regime, where $a$ is a parameter given as $a = \frac{\zeta[3]}{2 \zeta[2]} \simeq 0.365381$. We also present a concrete global construction using a $K3$-fibred CY threefold with $h^{1,1} =3$ which shares many of its properties with those of the standard toroidal case, and subsequently generates the appropriate corrections needed to fix all the three K\"ahler moduli at the perturbative level. We further discuss whether de Sitter vacua can be  ensured  through appropriate contributions of uplifting terms in the effective potential.
}

\keywords{String compactifications, de-Sitter Vacua, String loop corrections}
\begin{document}
\makeatletter
\let\old@fpheader\@fpheader
\renewcommand{\@fpheader}{\old@fpheader\hfill
arXiv:2203.03362}
\makeatother

\maketitle

\bigskip


\section{Introduction}
\label{sec_intro}

Currently, the fundamental problem of moduli stabilisation and the quest for de Sitter vacua  are  a subject of intensive research  activity  in string theory.  
 Despite the continuous efforts over the last couple of decades, both issues remain open  to this day.
Possible solutions, if they exist at all~\footnote{The recent literature is vast. For  comprehensive analyses and related work on these issues see the reviews~\cite{Danielsson:2018ztv,Palti:2019pca}.}, are sought beyond the classical level and are based on quantum corrections  which  modify   the K\"ahler potential
and the superpotential.  During the last few decades, 
a broad spectrum  of perturbative and non-perturbative contributions have been implemented to confront  these issues. Amongst the leading early proposals on this issue are the KKLT construction~\cite{Kachru:2003aw} and the LARGE volume scenario (LVS)~\cite{Balasubramanian:2005zx}. In these scenarios, stabilisation of the K\"ahler moduli is 
based on non-perturbative corrections~\cite{Witten:1996bn} in the superpotential while $\overline{D3}$ contributions \cite{Kachru:2003aw} or D-terms~\cite{Burgess:2003ic} are introduced to uplift the AdS vacuum to a de Sitter (dS) space. 
For the same reasons, the study of non-perturbative effects has been in the center of (K\"ahler) moduli stabilisation in the type IIB orientifold framework and several (new) mechanisms have been proposed in the meantime in order to induce such corrections in the effective four dimensional theory \cite{Blumenhagen:2009qh, Cvetic:2007sj, Bobkov:2010rf,Bianchi:2011qh}. However, given the fact that these non-perturbative effects have been proposed in the 4D effective theory and their higher dimensional origin has not been understood or clear enough yet, there have been observations regarding some incompatibilities while building realistic models, especially when the open string moduli are involved, e.g. see \cite{Blumenhagen:2007sm, Cvetic:2012ts}. Moreover, there are some arguments in recent literature~\cite{Carta:2019rhx,Kim:2022jvv}, which cast further doubts whether non-perturbative effects usually implemented in the model building are generic and genuine enough to play the decisive r\^ole which they have been attributed to. It is therefore reasonable to contemplate the idea  whether an elegant and viable solution could be achieved only with a minimum number of robust and well defined ingredients.   From this perspective, it would be  of particular interest to investigate whether  a successful  outcome can emerge by incorporating only perturbative quantum corrections. 
Therefore, in this work we focus on a minimal set of quantum  contributions which suffice to break the no-scale structure of the K\"ahler potential and are  more or less model independent.

Various sources of perturbative quantum corrections have been investigated over the last couple of decades.  We first mention the $(\alpha^\prime)^3$ corrections \cite{Becker:2002nn} which are proportional to the Euler characteristic of the internal manifold and lead to a non-vanishing scalar potential for the K\"ahler moduli. 
Also, the r\^ole of  string loop effects in the presence of D-branes and O-planes for large volume compactifications of toroidal orientifolds and other related issues have been extensively discussed in references~\cite{Berg:2005ja,Berg:2007wt,Haack:2018ufg, Cicoli:2007xp}. 
Furthermore there are one-loop logarithmic corrections   in the K\"ahler potential~\cite{Antoniadis:2018hqy,Antoniadis:2019rkh} stemming  from  configurations of  D7 brane stacks  and a novel four-dimensional Einstein-Hilbert term (localised within the six-dimensional internal space) generated from higher derivative terms in the ten-dimensional string effective action~\cite{Antoniadis:1997eg,Green:1997di,Kiritsis:1997em,Antoniadis:2002tr}. Within this framework, de Sitter minima can be achieved by virtue of  an `uplift' contribution emerging from D-terms  associated with the universal $U(1)$ factors of the  D7-stacks. In addition to the two-derivative ($F^2$) scalar potential effects arising from the K\"ahler potential and the superpotential, there have been a new class of higher derivative contributions found at the $F^4$-order~\cite{Ciupke:2015msa}. These contributions are also perturbative in nature, and appear with slightly more suppressed powers in terms of the overall volume of the internal manifold. Let us mention here that (most of) these aforesaid perturbative corrections are invoked by considering the reduction from a higher dimensional term, and in that sense their parental origin could be thought of being better understood. In this regard, using higher dimensional symmetries in the $F$-theory context, a systematic analysis of the volume/dilaton dependencies of several $\alpha^\prime$ as well as $g_s$ corrections have been presented in \cite{Cicoli:2021rub}. Moreover, let us add that these perturbative corrections we mentioned are quite generic in the sense that they can be (mostly) present in arbitrary CY orientifold models, unlike the non-perturbative effects which demand some very specific conditions in order to contribute to the superpotential, e.g. the unit arithmetic genus condition of \cite{Witten:1996bn}, or an ``appropriate" choice of fluxes to ``rigidify" the non-rigid divisors \cite{Bianchi:2011qh} or about finding an ample divisor which is rigid as well \cite{Bobkov:2010rf, Carta:2022web}.
 
In  the present work, taking into account the above mentioned perturbative ingredients, a systematically analytic treatment is performed aiming to obtain a closed simple formula for the scalar potential.  The analysis focuses only on the minimum radiative corrections which are necessary to guarantee the existence of  (anti-) de Sitter vacua in the large volume regime. In this approach, of primary importance are
perturbative $(\alpha^\prime)^3$ as well as string loop effects, especially the one-loop logarithmic corrections, whereas non-perturbative effects, for the reasons mentioned above, are assumed to be absent in explicit Calabi Yau orientifold constructions.  The moduli space K\"ahler metric and its inverse metric are constructed for a combination of $\alpha^\prime$ and the string-loop effects using the closed string chiral variables, namely the axio-dilaton ($S$), the complexified K\"ahler moduli $(T_\alpha)$ and the complex structure moduli $(U^i)$ from which a simple analytic form for the scalar potential is obtained while various limiting cases are considered. In addition, D-terms or $\overline{D3}$ contributions must be introduced to uplift the vacuum to a de Sitter space. In particular 
the expansions in $\alpha'$ and the logarithmic correction are presented and are found to be   in accordance with previous works.
A concise description of the model and the quantum corrections implemented is given in  section \ref{sec_no-scale-breaking}. In incorporating  the  perturbative corrections into the K\"ahler potential we recall  that type IIB string theory admits the discrete $SL(2, Z)$ symmetry  which  implies invariance of the resulting effective theory under some subgroup  $\Gamma_s\subset SL(2, Z)$. This fact motivates us to write the $\alpha'$ corrections in terms of the Eisenstein series $E_{3/2}$. Then, some of the quantum corrections appear in different powers of the $g_s$ expansion. 
In  section \ref{sec_master-formula}, a simple formula of 
the effective potential is derived which includes $\alpha'$ and logarithmic string loop corrections.  Various limiting cases are considered in the large volume regime. A generic formula of 
the scalar potential including also the sub-leading terms and higher derivative $F^4$ contributions is given in section \ref{sec_gen-formula}. In section \ref{sec_global-model} we present a concrete CY orientifold model which shares many of its properties with the standard six-torus orientifold case, and hence gets directly applicable to the current scenario. In section \ref{sec_moduli-stabilization},  the problem of moduli stabilisation is discussed in detail using the concrete specific data of the global CY orientifold, along with exploring numerical models with de Sitter vacua. 
Our  conclusions are presented in section \ref{sec_conclusions}.


\section{No-scale breaking through string loops}
\label{sec_no-scale-breaking}

The type IIB K\"ahler potential receives two kinds of corrections at the perturbative level; one arising from the $\alpha'$ series of higher derivatives effects  and the other one is induced through the string-loop $(g_s)$ corrections. Using appropriate chiral variables, a generic form for the K\"ahler potential incorporating (some of) the perturbative $\alpha^\prime$ and $g_s$ corrections can be written as the sum of two terms motivated by their underlying ${\cal N}=2$ special K\"ahler and quaternionic structure:
\bea
\label{eq:K}
& & {\cal K} = K_{cs} + K\,,
\eea
where:
\bea
\label{eq:defK}
& & K_{cs} = -\ln\left({\rm i}\int_X \Omega_3\wedge{\bar\Omega_3}\right) \qquad\text{and}\qquad K = - \ln\left(-{\rm i}(S-\ov{S})\right) -2\ln{\cal Y}\,.
\eea
Furthermore, various sub-leading corrections to the overall volume ${\cal V}$ can be encoded in ${\cal Y}$ expressed as ${\cal Y} = {\cal Y}_0 + {\cal Y}_1$ and defined in the following manner, 
\bea
& & {\cal Y}_0 = \vo +  \frac{\xi}{2} \, e^{-\frac{3}{2} \phi} = \vo + \frac{\xi}{2}\, \left(\frac{S-\ov{S}}{2\,{\rm i}}\right)^{3/2} \,, \nonumber\\
& & {\cal Y}_1 = e^{\frac{1}{2} \phi}\, f({\cal V}) = \left(\frac{S-\ov{S}}{2\,{\rm i}}\right)^{-1/2} f({\cal V})\,,
\label{eq:defY}
\eea
where $\vo$ is the tree-level CY volume $\vo = \frac16 \,{k_{\alpha\beta\gamma} \, t^\alpha\, t^\beta \, t^\gamma}$ in the Einstein frame and $\xi$ is proportional to the CY Euler characteristic $\chi$ such that $\xi = -\frac{\zeta(3)\, \chi(X)}{2\,(2\pi)^3}$. Here ${\cal Y}_0$ denotes the $\alpha'$ corrected CY volume \cite{Becker:2002nn} which is still at the `tree'-level in the string-loop series, while ${\cal Y}_1$ denotes the one-loop correction which can generically have dependence on the overall volume ${\cal V}$ as suggested in \cite{Antoniadis:2018hqy}. In fact, after including the $SL(2,{\mathbb Z})$ completion of the  $\alpha'$ corrections of \cite{Becker:2002nn} one gets non-holomorphic Eisenstein series  $E_{3/2}(S, \ov{S})$ defined as \cite{Grimm:2007xm},
\bea
& & E_{3/2}({S}, \ov{S}) = \sum_{(p,q) \neq (0,0)} \, \frac{\left({S} - \ov{S}\right)^{\frac{3}{2}}}{(2\, i)^{\frac{3}{2}} \, |p + q \, {S} |^{3}},
\eea
which in the weak coupling limit includes a constant perturbative term at one-loop order as can be seen from the expansion below,
\bea
\label{eq:Eisenstein-exp}
& & \hskip-1.5cm E_{3/2}({S}, \ov{S})  = 2\, \zeta[3] \, \left(\frac{{S} - \ov{S}}{2\,i}\right)^{3/2} + 4 \, \left(\frac{{S} - \ov{S}}{2\,i}\right)^{-1/2}\, \zeta[2] + \left(\frac{{S} - \ov{S}}{2\,i}\right)^{1/2} \, {\cal O}(e^{-2\pi s}),
\eea
where the first term corresponds to the BBHL pieces while the second term is proportional to $s^{-1/2}$ where $s = Re(S) =  g_s^{-1}$ given that string coupling is defined through $g_s = e^{\langle \phi \rangle}$. The last term corresponds to non-perturbative string-loop effects which we ignore for the current work. Combining the effects of \cite{Antoniadis:2018hqy} along with the above mentioned one-loop piece suggests the following form for the function $f({\cal V})$,
\bea
\label{eq:fvol}
& & f({\cal V}) = \sigma + \eta \, \ln{\cal V}\,,
\eea
where we have introduced two parameters $\sigma$ and $\eta$ which do not depend on any moduli. Moreover, let us note that the choice of implicit function $f({\cal V})$ in Eq. (\ref{eq:defY}) is well consistent with some more generic situations given that it depends only on the Einstein frame volume ${\cal V}$ which does not transform under the $SL(2,{\mathbb Z})$ transformations. Subsequently we have the following Ansatz for the shifted volume ${\cal Y}$ appearing in the K\"ahler potential in Eq.~(\ref{eq:defK}),
\bea
\label{eq:defY-simp}
& & {\cal Y} = \vo +  \frac{\xi}{2} \, e^{-\frac{3}{2} \phi} + e^{\frac{1}{2} \phi}\, \left(\sigma + \eta \, \ln{\cal V}\right).
\eea
Comparing the BBHL and 1-loop terms arising from the expansion of the Eisenstein series in Eq.~(\ref{eq:Eisenstein-exp}), along with the logarithmic loop corrections computed in  \cite{Antoniadis:2018hqy,Antoniadis:2018ngr,Antoniadis:2019doc,Antoniadis:2019rkh,Antoniadis:2020ryh,Antoniadis:2020stf} one can have the following correlations among the various coefficients, namely $\xi$, $\sigma$ and $\eta$,
\bea
\label{eq:def-xi-eta}
& & \hskip-1cm  \xi = - \frac{\chi(CY)\, \zeta[3]}{2(2\pi)^3}~, \qquad \sigma  = - \frac{\chi(CY)\, \zeta[2]}{2(2\pi)^3} = - \, \eta, \qquad \frac{\xi}{\eta} = -\frac{\zeta[3]}{\zeta[2]} \\
& & \hskip-1cm \hat\xi = \frac{\xi}{g_s^{3/2}}~, \qquad \hat\eta = g_s^{1/2}\, \eta~, \qquad \qquad \frac{\hat\xi}{\hat\eta} = -\frac{\zeta[3]}{\zeta[2]\,g_s^2}~. \nonumber
\eea

\subsection{Computation of the K\"ahler metric and its inverse}
We are using the following definitions of the chiral variables 
\bea
\label{eq:chiral-variables}
& & U^i = v^i - i\, u^i, \qquad S = c_0 + i\, e^{-\phi}, \qquad T_\alpha = c_\alpha - i\, \tau_\alpha~,
\eea
where $\phi$ is the dilaton, $u^i$'s are the complex structure saxions, and $\tau_\alpha$'s are the Einstein frame four-cycle volume moduli defined as $\tau_\alpha = \partial_{t^\alpha} {\cal V} = \frac12 k_{\alpha\beta\gamma} t^\beta t^\gamma$. In addition, the $C_0$ and $C_\alpha$'s are universal RR axion, RR four-form axions respectively while the complex structure axions are denoted as $v^i$. Here the indices $\{i, \alpha\}$ are such that $i \in h^{2,1}_-(CY/{\cal O})$ while $\alpha \in h^{1,1}_+(CY/{\cal O})$. Moreover, we assume that $h^{1,1} = h^{1,1}_+$ for simplicity, and hence there are no so-called odd-moduli $G^a$ which are present in our analysis, and we refer the interested readers to \cite{Cicoli:2021tzt}. Using the K\"ahler potential pieces  given in Eqs. (\ref{eq:defK})-(\ref{eq:defY}), we get the following useful relations in the absence of any odd-moduli $G^a$,
\bea
\label{eq:der-conds1}
& & \partial_S t^\alpha = 0 = \partial_{\ov{S}} t^\alpha, \qquad  \partial_{S} {\cal V} = 0 =  \partial_{\ov{S}} {\cal V}, \qquad \partial_{S} \hat\xi = - \frac{3\,i\,\hat\xi}{4\, s} = - \partial_{\ov{S}} \hat\xi, \\
& & \partial_{T_\beta} t^\alpha = \frac{i}{2}\,k^{\alpha\beta} = - \partial_{\ov{T}_\beta} t^\alpha, \qquad \partial_{T_\alpha} {\cal V} = \frac{i}{4}\,t^{\alpha} = - \partial_{\ov{T}_\alpha} {\cal V}, \qquad \partial_{T_\alpha} \hat\xi = 0 = \partial_{\ov{T}_\alpha} \hat\xi, \nonumber\\
& & \partial_{S} {\cal Y}_0 = - \frac{3\,i\,\hat\xi}{8\, s} = - \partial_{\ov S} {\cal Y}_0, \qquad  \partial_{S} {\cal Y}_1 = \frac{i}{4\, s} \, {\cal Y}_1 = -\partial_{\ov S} {\cal Y}_1, \nonumber\\
& & \partial_{T_\alpha} {\cal Y}_0 = \frac{i}{4}\,t^{\alpha} = -\partial_{\ov{T}_\alpha} {\cal Y}_0, \qquad \partial_{T_\alpha} {\cal Y}_1 = \frac{i}{4}\,t^{\alpha}\, s^{-\frac{1}{2}} \frac{\partial f}{\partial {\cal V}} = \frac{i}{4}\,t^{\alpha}\, \frac{\partial {\cal Y}_1}{\partial {\cal V}} = - \partial_{\ov{T}_\alpha} {\cal Y}_1, \nonumber
\eea
where we have introduced shorthand notation such as $k_\alpha t^\alpha = k_{\alpha\beta} t^\alpha t^\beta = k_{\alpha\beta\gamma} t^\alpha t^\beta t^\gamma = 6 {\cal V}$ and $k^{\alpha\beta} = (k_{\alpha\beta\gamma} t^\gamma)^{-1}$ which subsequently satisfies an identity: $k^{\alpha\beta} k_{\beta} = t^\alpha$. As and when needed we have also used $s = e^{-\phi}$ and $\hat\xi\equiv \xi \, s^{3/2}$. This also results in the following relation which can be directly used at various intermediate steps while computing the K\"ahler derivatives and the K\"ahler metric,
\bea
\label{eq:der-conds2}
& & \hskip-0cm \partial_S {\cal Y} = - \frac{i}{4\, s} \left(\frac{3\,\hat\xi}{2} -\, {\cal Y}_1 \right) = - \partial_{\ov S} {\cal Y}, \\
& & \partial_{T_\alpha} {\cal Y} = \frac{i}{4}\,t^{\alpha} \left(1 + \frac{\partial {\cal Y}_1}{\partial {\cal V}} \right) = - \partial_{\ov{T}_\alpha} {\cal Y}. \nonumber
\eea
One can observe that $\partial_S {\cal Y}$ does not have any tree-level piece and can receive non-zero contributions only from the $\alpha^\prime$ and $g_s$ induced effects. Subsequently, the K\"ahler derivatives for the quaternionic sector of the K\"ahler potential ($K$) given in Eq. (\ref{eq:defK}) can be generically expressed as below,
\bea
\label{eq:derK}
& & K_S = \frac{{\rm i}}{2 \,s }\left(1 + \frac{3\, \hat\xi}{2\,{\cal Y}} - \frac{{\cal Y}_1}{{\cal Y}}\right) = \frac{{\rm i}}{2 \,s \, {\cal Y}}\left({\cal V} + 2 \, \hat\xi \right) = - K_{\ov{S}}, \\
& & K_{T_\alpha} = -\frac{{\rm i} \, t^\alpha}{2\, {\cal Y}} \left( 1 + s^{-\frac{1}{2}}\, \frac{\partial f}{\partial{\cal V}} \right) = -\frac{{\rm i} \, t^\alpha}{2\, {\cal Y}} \left( 1 +  \frac{\partial {\cal Y}_1}{\partial{\cal V}} \right) = - K_{\ov{T}_\alpha}\,. \nonumber
\eea
Using these derivatives, the various K\"ahler metric components are found to be:
\bea
\label{eq:Kij}
& & K_{S \ov{S}} = \frac{1}{8\,s^2\, {\cal Y}^2}\, \left({\cal V} ({\cal Y} + {\cal V}) - 4 \hat\xi ({\cal Y} - {\cal V}) + 4 \hat\xi^2\right), \\
& & K_{T_\alpha\, \ov{S}} = -\frac{t^\alpha}{8\, s\, {\cal Y}^2} \left(\frac{3}{2}\, \hat\xi - s^{-\frac{1}{2}}\, f + \, s^{-\frac{1}{2}} \, ({\cal V} + 2\hat\xi) \frac{\partial f}{\partial {\cal V}} \right)= K_{S\,\ov{T}_\alpha}, \nonumber\\
& & K_{T_\alpha \, \ov{T}_\beta} = \frac{9\, {\cal G}^{\alpha\beta}}{4\, {\cal Y}^2} \left(1 + s^{-\frac{1}{2}} \, \frac{\partial f}{\partial {\cal V}} \right) - s^{-\frac{1}{2}} \, \frac{t^\alpha t^\beta}{8\, {\cal Y}} \, \frac{\partial^2 f}{\partial {\cal V}^2} , \nonumber
\eea
where, using our shorthand notation, the metric ${\cal G}$  and its inverse ${\cal G}^{-1}$, can be given by the following expressions,
\be
\label{eq:genMetrices}
\frac{{\cal G}_{\alpha\beta}}{36} = \frac{k_\alpha\,k_\beta}{4{\cal Y}\, (6\vo - 2\, {\cal Y})} -\frac{k_{\alpha\beta}}{4\, {\cal Y}} \qquad\text{and}\qquad 36\,{\cal G}^{\alpha\beta} = 2 \, t^\alpha \, t^\beta -4\, {\cal Y} \, k^{\alpha\beta}. 
\ee 
In fact, the true interpretation of the moduli space metric should be consider to be the one which is corrected by all the $\alpha^\prime$ and $g_s$ effects in a collective manner. Now, to facilitate an easier inversion of the moduli space metric we can express it in the following formulation,
\bea
\label{eq:simpKij-1}
& & \hskip-1cm K_{S \ov{S}} = {\cal P}_1, \qquad K_{T_\alpha\, \ov{S}} = t^\alpha\, {\cal P}_2 = K_{S\,\ov{T}_\alpha}, \qquad  K_{T_\alpha \, \ov{T}_\beta} = (t^\alpha \, t^\beta)\, {\cal P}_3 \, - k^{\alpha\beta}\, {\cal P}_4~,
\eea
where the four functions ${\cal P}_1, {\cal P}_2, {\cal P}_3$ and ${\cal P}_4$ are collected as below
\bea
\label{eq:Pis}
& & {\cal P}_1 = \frac{1}{8\,s^2\, {\cal Y}^2}\, \left({\cal V} ({\cal Y} + {\cal V}) - 4 \hat\xi ({\cal Y} - {\cal V}) + 4 \hat\xi^2\right), \\
& & {\cal P}_2 = -\frac{1}{8\, s\, {\cal Y}^2} \left(\frac{3}{2}\, \hat\xi - s^{-\frac{1}{2}}\, f + \, s^{-\frac{1}{2}} \, ({\cal V} + 2\hat\xi) \frac{\partial f}{\partial {\cal V}} \right), \nonumber\\
& & {\cal P}_3 = \frac{1}{8\, {\cal Y}^2} \left(1 + s^{-\frac{1}{2}} \, \frac{\partial f}{\partial {\cal V}} -  {\cal Y} \,  s^{-\frac{1}{2}} \, \frac{\partial^2 f}{\partial {\cal V}^2} \right), \nonumber\\
& & {\cal P}_4 = \frac{1}{4\, {\cal Y}} \left(1 + s^{-\frac{1}{2}} \, \frac{\partial f}{\partial {\cal V}} \right). \nonumber
\eea
Actually rewriting the K\"ahler metric in the above formulation admits the inverse K\"ahler metric of the following form,
\bea
\label{eq:simpinvKij-1}
& & \hskip-1cm  K^{S \ov{S}} = \tilde{\cal P}_1, \qquad K^{T_\alpha\, \ov{S}} = k_\alpha\, \tilde{\cal P}_2 = K^{S\,\ov{T}_\alpha}, \qquad  K^{T_\alpha \, \ov{T}_\beta} = (k_\alpha \, k_\beta)\, \tilde{\cal P}_3 \, - k_{\alpha\beta}\, \tilde{\cal P}_4~,
\eea
where the four new functions $\tilde{\cal P}_1, \tilde{\cal P}_2, \tilde{\cal P}_3$ and $\tilde{\cal P}_4$ are to be determined by demanding $K_{A\ov{B}} K^{\ov{B} C} = \delta_A{}^B$. Now using short hand notations/definitions and identities such as $k_\alpha t^\alpha = 6 {\cal V}, \, k_{\alpha\beta} t^\beta = k_\alpha$ and $k^{\alpha\beta} k_\beta = t^\alpha$, one can show that $K_{A\ov{B}} K^{\ov{B} C} = \delta_A{}^B$ results in the following set of constraints,
\bea
\label{eq:inversion-Pi-tPi}
& & {\cal P}_1 \tilde{\cal P}_1 + 6 {\cal V} {\cal P}_2 \tilde{\cal P}_2 = 1, \qquad {\cal P}_1 \tilde{\cal P}_2 +  {\cal P}_2 (6 {\cal V} \tilde{\cal P}_3 - \tilde{\cal P}_4) = 0, \\
& & {\cal P}_2 \tilde{\cal P}_2 + 6 {\cal V} {\cal P}_3 \tilde{\cal P}_3 - {\cal P}_4 \tilde{\cal P}_3 - {\cal P}_3 \tilde{\cal P}_4 = 0, \qquad {\cal P}_4 \tilde{\cal P}_4 = 1. \nonumber
\eea
Now,  this reduces the task of inverting the complicated K\"ahler metric into solving a set of four quadratic equations in four variables. Subsequently, the functions $\tilde{\cal P}_1, \tilde{\cal P}_2, \tilde{\cal P}_3$ and $\tilde{\cal P}_4$ are given as below,
\bea
\label{eq:tildePis}
& & \tilde{\cal P}_1 = \frac{{\cal P}_4-6 {\cal P}_3 {\cal V}}{{\cal P}_1 {\cal P}_4 + 6 {\cal P}_2^2 {\cal V}-6 {\cal P}_1 {\cal P}_3 {\cal V}}\,, \\
& & \tilde{\cal P}_2 = \frac{{\cal P}_2}{{\cal P}_1 {\cal P}_4 + 6 {\cal P}_2^2 {\cal V}-6 {\cal P}_1 {\cal P}_3 {\cal V}}\,, \nonumber\\
& & \tilde{\cal P}_3 = \frac{{\cal P}_2^2-{\cal P}_1 {\cal P}_3}{{\cal P}_4\left({\cal P}_1 {\cal P}_4 + 6 {\cal P}_2^2 {\cal V}-6 {\cal P}_1 {\cal P}_3 {\cal V}\right)}\,, \nonumber\\
& & \tilde{\cal P}_4 = ({\cal P}_4)^{-1}. \nonumber
\eea
Let us mention that it does not appear to be illuminating to give explicit generic expressions for $\tilde{\cal P}_i$'s in terms of the K\"ahler potential ingredients (such as ${\cal V}$, $\hat\xi$ and $f({\cal V}$)) which results in quite lengthy and complicated expressions, and can nevertheless be directly read-off from Eqs. (\ref{eq:Pis})-(\ref{eq:tildePis}). However it is worth to reproduce the known results as a limiting case of our general formulae. For example, in the absence of both of the $\alpha^\prime$ as well as $g_s$ corrections, the tree level expressions for ${\cal P}_i$ and $\tilde{\cal P}_i$ are given as below \cite{Grimm:2004uq},
\bea
\label{eq:Pi-tPi-tree}
& & {\cal P}_1 = \frac{1}{4\,s^2}, \qquad {\cal P}_2 = 0, \qquad {\cal P}_3 = \frac{1}{8\, {\cal V}^2}, \qquad {\cal P}_4 = \frac{1}{4\,{\cal V}}, \\
& & \tilde{\cal P}_1 = 4\, s^2, \qquad \tilde{\cal P}_2 = 0, \qquad \tilde{\cal P}_3 = 1, \qquad \tilde{\cal P}_4 = 4 {\cal V}. \nonumber
\eea
Similarly for the case of BBHL's $\alpha^\prime$ corrections \cite{Becker:2002nn} being included, we have the following simplified expressions for ${\cal P}_i$ and $\tilde{\cal P}_i$ which matches with those claimed in \cite{Bobkov:2004cy,AbdusSalam:2020ywo,Cicoli:2021tzt},
\bea
\label{eq:Pi-tPi-BBHL}
& & {\cal P}_1 = \frac{4{\cal V}^2 + \hat\xi \left({\cal V}+4 \hat\xi \right)}{16\,s^2\, ({\cal V}+\frac{1}{2}\hat{\xi})^2}, \qquad {\cal P}_2 = - \frac{3\, \hat\xi}{16\, s\,({\cal V}+\frac{1}{2}\hat{\xi})^2}, \\
& & {\cal P}_3 = \frac{1}{8\,({\cal V}+\frac{1}{2}\hat{\xi})^2}, \qquad {\cal P}_4 = \frac{1}{4\,({\cal V}+\frac{1}{2}\hat{\xi})}, \nonumber\\
& & \tilde{\cal P}_1 = \frac{s^2\,(4 \,{\cal V}-\hat{\xi})}{(\vo-\hat\xi)}, \quad \tilde{\cal P}_2 = \frac{3\, s\,\hat{\xi}}{2\,({\cal V}-\hat{\xi})}, \quad \tilde{\cal P}_3 = \frac{4 \,{\cal V}-\hat{\xi}}{4\,(\vo-\hat\xi)}, \quad \tilde{\cal P}_4 = 4 \left({\cal V}+\frac{1}{2}\hat{\xi}\right). \nonumber
\eea
In \cite{Cicoli:2021tzt} it was observed that the form of $\tilde{\cal P}_1$ remains the same even after including the odd-moduli, despite a complicated mixing with new terms in the generalized version of ${\cal P}_1$ in the corresponding  $K_{S, \ov{S}}$ component of the K\"ahler metric.

\subsection{Some useful no-scale breaking identities}
Considering the explicit form of the K\"ahler derivatives inEq.~(\ref{eq:derK}) along with the various components of the inverse K\"ahler metric in Eq.~(\ref{eq:simpinvKij-1}) supplemented by Eq.~(\ref{eq:Pis}) and Eq.~(\ref{eq:tildePis}), we find the following useful simplified relations:
\bea
\label{eq:identities1}
K_S\,K^{S\ov{S}} &=& \frac{{\rm i ({\cal V} + 2 \, \hat\xi )}}{2 \,s \, {\cal Y}} \tilde{\cal P}_1 = -\, K^{S \ov{S}} \, K_{\ov{S}},   \\
K_S \, K^{{S} \ov{T}_\alpha}  &=&\frac{{\rm i ({\cal V} + 2 \, \hat\xi )}}{2 \,s \, {\cal Y}} \, k_\alpha\,\tilde{\cal P}_2 = -\, K^{{T_\alpha} \ov{S}} \, K_{\ov{S}},   \nn \\
K_{T_\alpha} \, K^{T_\alpha\,\ov{S}} &=& -\frac{3\, {\rm i} \, {\cal V}}{{\cal Y}} \left( 1 +  \frac{\partial {\cal Y}_1}{\partial{\cal V}} \right) \, \tilde{\cal P}_2= -\, K^{S\,\ov{T}_\alpha} \, K_{\ov{T}_\alpha},  \nn \\
K_{T_\alpha} \, K^{{T_\alpha} \ov{T}_\beta}  &=&  -\frac{{\rm i} }{2\, {\cal Y}} \left( 1 +  \frac{\partial {\cal Y}_1}{\partial{\cal V}} \right) \left(6 {\cal V} \tilde{\cal P}_3 - \tilde{\cal P}_4 \right) \, k_\beta= -\, K^{{T_\beta} \ov{T}_\alpha} \, K_{\ov{T}_\alpha}\,. \nn
\eea
In addition, we have the following relations:
\bea
\label{eq:identities2}
K_S \, K^{S \ov{S}} \, K_{\ov{S}} &=& \frac{({\cal V} + 2 \, \hat\xi)^2}{4 \,s^2 \, {\cal Y}^2} \tilde{\cal P}_1 \,, \\
K_S \, K^{{S} \ov{T}_\alpha} \, K_{\ov{T}_\alpha} &=& -\frac{3\, {\cal V}\,({\cal V} + 2 \, \hat\xi )}{2 \,s \, {\cal Y}^2} \, \left( 1 +  \frac{\partial {\cal Y}_1}{\partial{\cal V}} \right)\,\tilde{\cal P}_2 = K_{T_i} \, K^{{T_i} \ov{S}} \, K_{\ov{S}} \,, \nn \\
K_{T_\alpha} \, K^{{T_\alpha} \ov{T}_\beta} \, K_{\ov{T}_\beta} &=& \frac{3\, {\cal V} }{2\, {\cal Y}^2} \left(1 + \frac{\partial {\cal Y}_1}{\partial{\cal V}}\right)^2 \left(6 {\cal V} \tilde{\cal P}_3 - \tilde{\cal P}_4 \right) \,. \nn
\eea
Moreover, one can find the following useful relations using Eqs~(\ref{eq:identities1})-(\ref{eq:identities2}),
\begin{align}
\label{eq:identities3}
{K}_A\, {K}^{{A} \ov{S}} &= \frac{{\rm i ({\cal V} + 2 \, \hat\xi )}}{2 \,s \, {\cal Y}} \tilde{\cal P}_1 -\frac{3\, {\rm i} \, {\cal V}}{{\cal Y}} \left( 1 +  \frac{\partial {\cal Y}_1}{\partial{\cal V}} \right) \, \tilde{\cal P}_2 = - {K}^{{S} \ov {B}} \, {K}_{\ov B} \,,\\
{K}_A\, {K}^{{A} \ov{T}_\alpha} &= k_\alpha\, \biggl[\frac{{\rm i ({\cal V} + 2 \, \hat\xi )}}{2 \,s \, {\cal Y}} \, \tilde{\cal P}_2 -\frac{{\rm i} }{2\, {\cal Y}} \left( 1 +  \frac{\partial {\cal Y}_1}{\partial{\cal V}} \right) \left(6 {\cal V} \tilde{\cal P}_3 - \tilde{\cal P}_4 \right)\biggr] = - {K}^{{T_\alpha} \ov {B}} \,  {K}_{\ov B} \,. \nonumber
\end{align}
In addition we have the following identitiy,
\bea
\label{eq:identities4}
& & K_A \, K^{A \ov{B}} \, K_{\ov{B}} = \frac{({\cal V} + 2 \, \hat\xi)^2}{4 \,s^2 \, {\cal Y}^2} \tilde{\cal P}_1 -\frac{3\, {\cal V}\,({\cal V} + 2 \, \hat\xi )}{ \,s \, {\cal Y}^2} \, \left( 1 +  \frac{\partial {\cal Y}_1}{\partial{\cal V}} \right)\,\tilde{\cal P}_2  \nn \\
& & \hskip1.5cm  + \frac{3\, {\cal V} }{2\, {\cal Y}^2} \left(1 + \frac{\partial {\cal Y}_1}{\partial{\cal V}}\right)^2 \left(6 {\cal V} \tilde{\cal P}_3 - \tilde{\cal P}_4 \right) \,. 
\eea
These identities can be directly used for deriving the generic formula for the scalar potential. 

Moreover, these expressions are given in full generality and it would be worth to present the particular limiting cases so that to understand and connect with the insights behind these identities. For example, when $\alpha'$ corrections are turned off, say via setting $\hat\xi=0$ along with the string-loop effects, using Eq. (\ref{eq:Pi-tPi-tree}) we get the following well-known tree-level results:
\bea
K_S \, K^{S\ov{S}} &=& 2\,{\rm i}\,s = -\, K^{S \ov{S}} \, K_{\ov{S}}, \quad K_S \, K^{{S} \ov{T}_\alpha} = 0 = K^{{T_\alpha} \ov{S}} \, K_{\ov{S}},  \\
K_{T_\alpha} \, K^{T_\alpha\,\ov{S}} &=& 0 = K^{S\,\ov{T}_\alpha} \, K_{\ov{T}_\alpha}, \quad K_{T_\alpha} \, K^{{T_\alpha} \ov{T}_\beta} = - \, {\rm i}\, k_\alpha = -\, K^{{T_\beta} \ov{T}_\alpha} \, K_{\ov{T}_\alpha}, \nn \\
K_S \, K^{S \ov{S}} \, K_{\ov{S}} &=& 1, \quad K_{S} \, K^{{S} \ov{T}_\alpha} \, K_{\ov{T}_\alpha} = 0, \quad K_{T_\alpha} \, K^{{T_\alpha} \ov{T}_\beta} \, K_{\ov{T}_\beta} = 3\,.\nn 
\eea
For the case of including the BBHL's $\alpha^\prime$ corrections, and in the absence of string-loop effects, the identities in Eq.~(\ref{eq:identities3}) and Eq.~(\ref{eq:identities4}) take the following simple form,
\bea
\label{eq:identities5}
& & {K}_A\, {K}^{{A} \ov{S}} = ({S} -\ov{S}) = - {K}^{{S} \ov {B}} \, {K}_{\ov B} \,,\\
& & {K}_A\, {K}^{{A} \ov{T}_\alpha} = (T_\alpha -\ov T_\alpha) = - {K}^{{T_\alpha} \ov {B}} \,  {K}_{\ov B} \,, \nn\\
& & K_A \, K^{A \ov{B}} \, K_{\ov{B}} = 4, \nn
\eea
where we have used explicit expressions of $\tilde{\cal P}_i$ as given in Eq.~(\ref{eq:Pi-tPi-BBHL}). In fact it was observed in \cite{AbdusSalam:2020ywo,Cicoli:2021tzt} that the above identities (\ref{eq:identities5}) which are usually well known to hold at the tree-level are promoted to hold even after including the BBHL's $\alpha^\prime$ corrections. In order to see some insights of adding the one-loop effects in the identities (\ref{eq:identities5}), let us consider $f({\cal V}) = \sigma$ where $\sigma$ is just some constant parameter which leads to the following modifications:
\bea
\label{eq:identities6}
& & {K}_A\, {K}^{{A} \ov{S}} = ({S} -\ov{S}) \left( 1 + \frac{2 \, e^{\frac12 \phi} \, \sigma}{{\cal Y} - 2 \,e^{\frac12 \phi} \, \sigma} \right)= - {K}^{{S} \ov {B}} \, {K}_{\ov B} \,,\\
& & {K}_A\, {K}^{{A} \ov{T}_\alpha} = (T_\alpha -\ov T_\alpha)\left( 1 - \frac{2 \, e^{\frac12 \phi} \, \sigma}{{\cal Y} - 2 \,e^{\frac12 \phi} \, \sigma} \right) = - {K}^{{T_\alpha} \ov {B}} \,  {K}_{\ov B} \,. \nn\\
& & K_A \, K^{A \ov{B}} \, K_{\ov{B}} = 4 + \frac{4 \, e^{\frac12 \phi} \, \sigma}{{\cal Y} - 2 \,e^{\frac12 \phi} \, \sigma}. \nonumber
\eea


\section{Combining BBHL and (logarithmic) string-loop effects}
\label{sec_master-formula}

In this section we present a generic formula for the scalar potential which includes $\alpha^\prime$ corrections of \cite{Becker:2002nn,Ciupke:2015msa} as well as (some of) the string-loop effects \cite{Antoniadis:2018hqy} via considering the so-called Gukov-Vafa-Witten's flux superpotential \cite{Gukov:1999ya}. We will subsequently use the master formula to read-off the scalar potentials for a set of Ans\"atze specific for the function $f({\cal V})$ as particular cases. Then we will present the moduli stabilisation and de Sitter realisation in one of the upcoming sections.

\subsection{Generic scalar potential}

The block diagonal nature of the total K\"ahler metric (and its inverse) with respect to the complex structure moduli sector and the remaining moduli sector admits the following splitting of contributions in the scalar potential,
\be
\label{eq:V_gen}
e^{- {\cal K}} \, V = {\cal K}^{{\cal A} \ov {\cal B}} \, (D_{\cal A} W) \, (D_{\ov {\cal B}} \ov{W}) -3 |W|^2 \equiv V_{\rm cs} + V_{\rm k}\,,
\ee
where:
\be
\label{eq:VcsVk}
V_{\rm cs} =  K_{\rm cs}^{i \ov {j}} \, (D_i W) \, (D_{\ov {j}} \ov{W}) \qquad \text{and}\qquad V_{\rm k} =  K^{{A} \ov {B}} \, (D_{A} W) \, (D_{\ov {B}} \ov{W}) -3 |W|^2\,.
\ee
Recall that the indices $(i, j)$ correspond to the complex structure moduli $U^i$ while the indices $(A,B)$ run over the remaining chiral variables $\{S, G^a, T_\alpha\}$ where $\alpha \in h^{1,1}_+(CY)$ and $a \in h^{1,1}_-(CY)$. However for our current purpose, we assume that the choice of the orientifold involution is such that the odd $(1,1)$-cohomology sector is trivial, and so there will be no odd moduli $G^a$ present in the current analysis. Assuming that the complex structure moduli and the axio-dilaton are fixed by supersymmetric $F$-flatness condition using the so-called Gukov-Vafa-Witten flux superpotential $W = W(S,U^i)$ we get,
\bea
\label{eq:susyDW}
&& D_i W = 0 = D_{\ov {i}} \ov{W}, \qquad D_{S} W = 0 = D_{\ov {S}} \ov{W}.
\eea
Subsequently, after fixing the complex structure moduli and the axion-dilaton by the leading order effects, the scalar potential  for the K\"ahler moduli can be generically given as below,
\bea
\label{eq:masterV}
& & \hskip-0.4cm V_{\alpha^\prime + {\rm log} \, g_s} =   e^{{\cal K}} \, |W|^2\, \left(K_{T_\alpha} \, K^{{T_\alpha} \ov{T}_\beta} \, K_{\ov{T}_\beta} -3 \,\right) \\
& & = e^{{\cal K}}\, |W|^2 \biggl[\frac{3\, {\cal V} }{2\, {\cal Y}^2} \left(1 + \frac{\partial {\cal Y}_1}{\partial{\cal V}}\right)^2 \left(6 {\cal V} \tilde{\cal P}_3 - \tilde{\cal P}_4 \right) - 3 \biggr]. \nonumber
\eea
As a warm-up to illustrate the utility of the master formula (\ref{eq:masterV}) let us quickly consider the tree-level case\footnote{Whenever we say tree level, we mean that both the $\alpha^\prime$ as well as the $g_s$ corrections are absent. For us, tree level should not be confused to be tree level in $g_s$ series only, as in that case BBHL corrections which are tree-level in string-loops can be allowed.}. From Eq. (\ref{eq:Pi-tPi-tree}) we read-off that $\tilde{\cal P}_3 = 1$ and $\tilde{\cal P}_4 = 4 {\cal V}$ while ${\cal Y}_1$ being a purely string-loop effect vanishes, along with neglecting the BBHL's $\alpha^\prime$ corrections leading to ${\cal Y} = {\cal V}$. With these pieces of information, one can immediately read-off from Eq. (\ref{eq:masterV}) that the scalar potential vanishes which is rooted in the so-called no-scale structure.

As a second example, let us consider the BBHL corrections without any string-loop effects. Subsequently, for this case using Eq. (\ref{eq:defY}) and Eq. (\ref{eq:Pi-tPi-BBHL}) we read-off the following details,
\bea
& & {\cal Y} =  \left({\cal V}+\frac{1}{2}\hat{\xi}\right), \quad {\cal Y}_1 = 0, \quad \quad \tilde{\cal P}_3 = \frac{4 \,{\cal V}-\hat{\xi}}{4\,(\vo-\hat\xi)}, \quad \tilde{\cal P}_4 = 4 \left({\cal V}+\frac{1}{2}\hat{\xi}\right)\,, \nonumber
\eea
which, using the master formula (\ref{eq:masterV}), recovers the well known BBHL piece given as below,
\bea
\label{eq:master2BBHL}
& & V_{\alpha^\prime} = e^{{\cal K}} \,  |W|^2\, \frac{3\, \hat{\xi }\left({\cal V}^2+7 \hat{\xi } \, {\cal V} + \, \hat{\xi }^2\right)}{4\,\left({\cal V}-\hat{\xi }\right) \left({\cal V}+\frac{1}{2}\hat{\xi}\right)^2}  \quad \xrightarrow[]{{\cal V} \to \infty} \quad \kappa \, \frac{3\, \hat\xi}{4\, {\cal V}^3} |W_0|^2,
\eea
where $\kappa = g_s/(8\pi)$, and $e^{K_{cs}} = 1$ which we set throughout the paper from now onwards. Here in the last step we have introduced an appropriately normalized flux superpotential parameter $W_0$ which is given as below \cite{Conlon:2006gv},
\begin{equation}
\label{eq:W0-value}
W = \sqrt{\frac{g_s}{8\,\pi}} \left\langle e^{\frac12 K_{cs}} \int_{\rm CY} \left(F_3 - S H_3\right) \wedge \Omega_3 \right\rangle = \sqrt{\frac{g_s}{8\,\pi}} \langle e^{\frac12 K_{cs}} \rangle \, W_0, 
\end{equation}
where $\Omega_3$ is the nowhere vanishing holomorphic three-form of the compactifying Calabi Yau threefold while $(F_3, H_3)$ denotes the ${\cal S}$-dual pair of RR and NS-NS three-form fluxes. We also note that the need for appropriately considering this overall factor (which a priory does not appear to play any role in the moduli dynamics) is the fact that we will be using not only the two-derivative contributions at order $F^2$ but also some higher derivative $F^4$-corrections, which leads to contributions with an overall factor $e^{2{\cal K}}\, |W|^4$.

\subsection{Approximating the scalar potential in weak coupling limit}
The very fact that the shifted volume in the K\"ahler potential appearing through ${\cal Y}$ involves an implicit function of the overall volume, ${\cal V}$,  in the form of $f({\cal V})$, through the one-loop piece ${\cal Y}_1$, and therefore it is not possible to make an explicit large volume expansion of the master formula (\ref{eq:masterV}) at this stage. However given that the $g_s$ dependencies are explicitly known and therefore it is indeed possible to make a weak coupling expansion which gives the following pieces,
\bea
\label{eq:masterVgen-simp}
& & V_{\alpha^\prime + {\rm log} \, g_s} =  \frac{12\, \kappa \, \hat{\xi }\left({\cal V}^2+7 \hat{\xi } \, {\cal V} + \, \hat{\xi }^2\right)}{\left(\, {\cal V}-\hat{\xi }\right) \left(2 \, {\cal V}+\hat{\xi }\right)^4} \,  |W_0|^2 \\
& & \hskip0.75cm + \frac{3\,\kappa\, \sqrt{g_s}}{\,\left({\cal V}-\hat{\xi }\right)^2 \left(2 \, {\cal V}+\hat{\xi }\right)^6} \,  |W_0|^2 \sum_{i=0}^7 q_i({\cal V}) \, {\cal V}^i + {\cal O}(g_s^2) + \cdots, \nonumber
\eea
where 
$q_i({\cal V})$'s are implicit functions depending on $f({\cal V})$ and can be given as below,
\bea
\label{eq:qis}
& & q_0({\cal V}) = 16 f \hat{\xi}^5, \\
& & q_1({\cal V}) = 160 f \hat{\xi}^4-16 \left(\frac{\partial f}{\partial {\cal V}}\right) \hat{\xi}^5, \nonumber\\
& & q_2({\cal V}) = 48 \left(\frac{\partial^2 f}{\partial {\cal V}^2}\right) \hat{\xi}^5+128 \left(\frac{\partial f}{\partial {\cal V}}\right) \hat{\xi}^4+184 f \hat{\xi}^3, \nonumber\\
& & q_3({\cal V}) = 120 \left(\frac{\partial^2 f}{\partial {\cal V}^2}\right) \hat{\xi}^4+161 \left(\frac{\partial f}{\partial {\cal V}}\right) \hat{\xi}^3-232 f \hat{\xi}^2, \nonumber\\
& & q_4({\cal V}) = 147 \left(\frac{\partial^2 f}{\partial {\cal V}^2}\right) \hat{\xi}^3+232 \left(\frac{\partial f}{\partial {\cal V}}\right) \hat{\xi}^2-160 f
\hat{\xi}, \nonumber\\
& & q_5({\cal V}) = 222 \left(\frac{\partial^2 f}{\partial {\cal V}^2}\right) \hat{\xi}^2+160 \left(\frac{\partial f}{\partial {\cal V}}\right) \hat{\xi}+32 f, \nonumber\\
& & q_6({\cal V}) = 96 \left(\frac{\partial^2 f}{\partial {\cal V}^2}\right) \hat{\xi}+64 \left(\frac{\partial f}{\partial {\cal V}}\right), \nonumber\\
& & q_7({\cal V}) = 96 \left(\frac{\partial^2 f}{\partial {\cal V}^2}\right).\nonumber
\eea
Note that, while making the weak coupling expansion we have considered ${\cal V}$ and $\hat\xi$ as Einstein frame quantities without pulling out the $g_s$ factors within them, relating to their respective string-frame expressions. The main motivation for us has been only to pull out the ``relative" factors for BBHL term appearing at $g_s^0$ order and the one-loop  effects at order $g_s^{1/2}$ as compared to BBHL while working with the Einstein frame ingredients.

Now let us make some observations about the form of the function $f({\cal V})$ which we have taken to be quite generic in our scalar potential formulation. The second (BBHL) piece of the simplified scalar potential (\ref{eq:masterVgen-simp}) can be seen to be of order ${\cal V}^{-3}$ while the third piece which results from string-loop effects involves two factors; the first of which corresponds to order ${\cal V}^{-8}$ multiplied by an implicit function $\left(q_i {\cal V}^i\right)$ which is a septic polynomial in ${\cal V}$ along with factors of derivatives and double derivatives of $f({\cal V})$. This analysis has a very interesting implication about the form of $f({\cal V})$ which is the fact that it cannot be a polynomial of ${\cal V}$ with positive powers as in that case the loop effects will start dominating the BBHL which is tree-level for string-loop series, and hence creating trouble for the whole perturbative notion of quantum corrections and their effective field theory description. To be more specific, let us consider
\bea
\label{eq:f-poly}
& & f({\cal V}) = \sum_{n \in {\mathbb Z}} f_n \, {\cal V}^n.
\eea
We subsequently find that the volume scaling for various pieces $q_i$'s are as follows:
\bea
& & q_0 \sim q_1 \sim q_2 \sim q_3 \sim q_4 \sim q_5 \sim {\cal V}^n,\quad q_6 \sim {\cal V}^{n-1}, \quad q_7 \sim {\cal V}^{n-2},
\eea
which shows that the volume factor in the string-loop pieces of (\ref{eq:masterVgen-simp}) scales in the following form,
\bea
{\cal V}^{-8} \times \left(\sum_{i=0}^7 q_i({\cal V})\, \, {\cal V}^i \right) \sim {\cal V}^{n-3}.
\eea
Now, given that the BBHL piece can be thought of a correction at tree-level in string-loops and hence one-loop effects should preferably not overtake it. This suggests that $n$ cannot be a positive number as
\bea
& & \frac{V_{g_s}}{V_{(\alpha^\prime)^3}} \leq 1 \quad \implies \quad n \leq 0.
\eea
This simple analysis suggests that the implicit function $f({\cal V})$ should not be a positive polynomial in ${\cal V}$, however polynomials with non-positive powers should be consistent with the validity of perturbative series expansion. In this regard, it is interesting to note that the $n = 0$ case corresponds to $f({\cal V}) = {\rm const.}$, which we have motivated by the $SL(2,{\mathbb Z})$ completion arguments of BBHL piece. In addition to that, a logarithmic nature of $f({\cal V})$ is always consistent with these arguments as its derivatives would be polynomials of ${\cal V}$ with negative powers. We will get back to the generic scalar potential regarding moduli stabilisation aspects later on.

\subsection{Analysing a set of string-loop scenarios}
In this section we will anayse a couple of Ans\"atze for the function $f({\cal V})$ responsible to induce the string-loop effects. 

\subsubsection*{Ansatz-1: $f({\cal V}) = \sigma$}
Let us investigate the effects of the simplest string-loop term in the K\"ahler potential by considering the function $f({\cal V}) = \sigma$ where $\sigma$ is some constant parameter. Also momentarily, let us switch-off the BBHL's $\alpha^\prime$ correction. Subsequently, the functions $\tilde{\cal P}_i$'s are given as,
\bea
\label{eq:Pi-tPi-loops1}
& & {\tilde{\cal P}_1} = \frac{2 s^{3/2} \left(2 s {\cal V}^2+\sqrt{s} \sigma  {\cal V} -\sigma ^2\right)}{{\cal V} \left(\sqrt{s}
   {\cal V}-\sigma \right)}, \qquad {\tilde{\cal P}_2} = - \frac{\sqrt{s} \sigma  \left(\sqrt{s} {\cal V}+\sigma \right)}{{\cal V} \left(\sqrt{s}
   {\cal V}-\sigma\right)}, \\
& & {\tilde{\cal P}_3} =  \frac{2 s {\cal V}^2+\sqrt{s} \sigma  {\cal V} -\sigma ^2}{2 s {\cal V}^2-2 \sqrt{s} \sigma 
   {\cal V}}, \qquad {\tilde{\cal P}_4} = 4 \left({\cal V} + \frac{\sigma }{\sqrt{s}}\right). \nonumber
\eea
Now, using our master formula (\ref{eq:masterV}) gives the scalar potential of the following form,
\bea
\label{eq:master2loops-constant}
& & V_{g_s}^{(1)} = e^{{\cal K}}\, |W|^2 \biggl[\frac{3 \sigma \left(\sqrt{s} {\cal V} + 2 \sigma\right)}{2 \left(s {\cal V}^2 - \sigma^2\right)} \biggr] \quad \xrightarrow[]{{\cal V} \to \infty} \quad \frac{3\,\kappa\, \sqrt{g_s} \, \sigma}{2\, {\cal V}^3} |W_0|^2.
\eea
Let us make an observation that the scalar potential arising from this type of string-loop effect results in a scalar potential contribution which is similar to the BBHL's $\alpha^\prime$ correction as given in Eq.~(\ref{eq:master2BBHL}) , though it has an additional string-loop suppression as expected, and can be seen from the large volume expansions leading to,
\bea
& & \frac{V_{g_s}^{(1)}}{V_{\alpha^\prime}} \simeq \frac{2\, \sqrt{g_s}\,\sigma}{\hat\xi} = g_s^2 \, \frac{2\, \sigma}{\xi},
\eea
where we have used $\hat\xi = g_s^{-3/2}\, \xi$ in the last step. Thus we can see clearly that the 1-loop effects are suppressed by a factor of $g_s^2$ as compared to the classical BBHL piece. This new term $V_{g_s}^{(1)}$ as given in Eq.~(\ref{eq:master2loops-constant}) may be useful for moduli stabilisation in some special circumstances. Moreover, if we include the BBHL contributions along with the string-loop terms having $f({\cal V}) = \sigma$, then Eq. (\ref{eq:master2loops-constant}) generalises to the following form,
\bea
& & \hskip-1cm V_{\alpha^\prime +g_s}^{(1)} =  e^{{\cal K}}\, |W|^2 \biggl[\frac{3 \left(\hat{\xi }^3 s+\hat{\xi } \left(s {\cal V}^2-8 \sqrt{s} \sigma  {\cal V} -4 \sigma ^2\right)+7 \hat{\xi }^2 s {\cal V}+2
   \sigma  {\cal V} \left(\sqrt{s} {\cal V}+2 \sigma \right)\right)}{\left({\cal V}-\hat{\xi }\right) \left(4
   s {\cal V}^2+4 \hat{\xi } s {\cal V} + \hat{\xi }^2 s-4 \sigma ^2\right)}  \biggr].
\eea

\subsubsection*{Ansatz-2: $f({\cal V}) = \sigma + \eta \ln{\cal V}$}
Now we investigate the effects of a bit less simple string-loop term in the K\"ahler potential by considering the function $f({\cal V}) = \sigma + \eta \ln{\cal V}$ where $\sigma$ and $\eta$ are some constant parameters. To begin with, let us again switch-off the BBHL's $\alpha^\prime$ correction. Subsequently, the simplified scalar potential can be given as below,
\bea
\label{eq:master2loops-Logvol}
& & V_{{\rm log} \, g_s} = \frac{e^{{\cal K}}\, |W|^2}{{\cal X}}\, \biggl[3 (-\eta +\sigma +\eta  \ln{\cal V}) \\
& & \hskip1cm \times \left(2 s {\cal V}^2+\sqrt{s} {\cal V} (4 \sigma -7 \eta )+2 \eta  \ln{\cal V} \left(2 \sqrt{s} {\cal V}-\eta \right) -2 \eta  \sigma\right) \biggr], \nonumber
\eea
where
\bea
& & {\cal X} = 2 \eta  \sigma ^2+4 s^{3/2} {\cal V}^3+10 \eta  s {\cal V}^2+\sqrt{s}
   {\cal V} \left(-3 \eta ^2+15 \eta  \sigma -4 \sigma ^2\right) \\
& & \hskip1cm +\eta  \ln{\cal V} \left(4 \eta  \sigma +\sqrt{s} {\cal V} (15 \eta
   -8 \sigma )+2 \eta  \ln{\cal V} \left(\eta -2 \sqrt{s} {\cal V}\right)\right). \nonumber
\eea
Taking the weak coupling limit of Eq.(\ref{eq:master2loops-Logvol}) we get
\bea
& & \hskip-1cm V_{{\rm log} \, g_s} \simeq \frac{3 \, \kappa \, |W_0|^2 (\hat\sigma -\hat\eta +\hat\eta  \ln{\cal V})}{2{\cal V}^3}-\frac{9 \, \kappa\, \hat\eta \, |W_0|^2 (\hat\sigma -\hat\eta +\hat\eta  \ln{\cal V})}{ {\cal V}^4} + \cdots,
\eea
where we have introduced the new parameters $\hat\sigma = \sqrt{g_s}\,\sigma$ and $\hat\eta = \sqrt{g_s}\,\eta$ similar to our $\alpha^\prime$ parameter earlier redefined as $\hat\xi = g_s^{-3/2}\xi$. Analogously, expanding the total scalar potential obtained from the master formula (\ref{eq:masterV}) in the presence of BBHL's $\alpha^\prime$ corrections, we get the following form of the leading order pieces,
\bea
\label{eq:pheno-potV1}
&& V_{\alpha^\prime +{\rm log} \, g_s} = \frac{12\, \kappa\,  |W_0|^2 \,  \hat{\xi }
   \left({\cal V}^2+7 \hat{\xi } {\cal V} + \hat{\xi }^2\right)}{\left({\cal V}-\hat{\xi }\right) \left(2
   {\cal V}+ \hat{\xi}\right)^4} + \frac{6\,\kappa\,  |W_0|^2}{\left({\cal V}-\hat{\xi }\right) \left(2 {\cal V} + \hat{\xi }\right)^6} \biggl[16 {\cal V}^4 (\hat\sigma -\hat\eta +\hat\eta  \ln{\cal V}) \nonumber\\
& & \hskip1.5cm -16 \,\hat{\xi } \,{\cal V}^3
   (4 \hat\sigma -\hat\eta +4 \hat\eta  \ln{\cal V})-3 \hat{\xi }^2 {\cal V}^2 (60 \hat\sigma -7 \hat\eta +60 \hat\eta  \ln{\cal V}) \nonumber\\
& & \hskip1.5cm  -4 \, \hat{\xi }^3 \, {\cal V} \,(22 \hat\sigma -7 \hat\eta +22\,\hat\eta  \ln{\cal V})-8 \hat{\xi }^4 (\hat\sigma -4 \hat\eta +\hat\eta  \ln{\cal V})\biggr] + {\cal O}(g_s^2)
\eea
Now we can also make another expansion for the large volume regime which leads to the following simplified form of the scalar potential,
\bea
\label{eq:pheno-potV2}
& & V_{\alpha^\prime +{\rm log} \, g_s}^{(1)} = \frac{3\, \kappa\, \hat\xi}{4\, {\cal V}^3}\, |W_0|^2 + \frac{3 \, \kappa\, (\hat\sigma -\hat\eta +\hat\eta  \ln{\cal V})}{2{\cal V}^3}\,|W_0|^2.
\eea
Note that the overall factors for the two terms in the string-loop effects are such that $\sigma = - \eta$, or equivalently $\hat\sigma = - \hat\eta$ as mentioned in Eq.~(\ref{eq:def-xi-eta}) which results in the following,
\bea
\label{eq:pheno-potV2}
& & V_{\alpha^\prime +{\rm log} \, g_s}^{(1)} = \frac{3\, \kappa\, \hat\xi}{4\, {\cal V}^3}\, |W_0|^2 - \frac{3 \, \kappa\, (2\,\hat\eta -\hat\eta  \ln{\cal V})}{2{\cal V}^3}\,|W_0|^2.
\eea
We will utilise this scalar potential for moduli stabilisation purpose in the next section.

\subsection{Perturbative LVS}
With the master formula (\ref{eq:masterV}) and the simplified versions derived so far, e.g. Eq.~(\ref{eq:pheno-potV2}), we are now in a position to perform the study of moduli stabilisation for the overall volume mode. For this purpose, the main idea is to fix the overall volume by the leading order ${\cal O}({\cal V}^{-3})$ terms arising from BBHL and logarithmic-loop effects, and then fix the remaining moduli by ${\cal O}({\cal V}^{-3-n})$ effects which will be a collective contributions induced from BBHL, string-loop effects as well as the $F^4$-terms as we will show later on. So we consider the simplified version of the scalar potential in Eq. (\ref{eq:pheno-potV2}) given as below,
\bea
\label{eq:pheno-potV3}
& & \hskip-1cm V_1 \simeq {\cal C}_1 \left(\frac{\hat\xi - 4\,\hat\eta + 2\,\hat\eta \, \ln{\cal V}}{{\cal V}^3}\right) + {\cal O}({\cal V}^{-4}) + \cdots \,, \quad {\cal C}_1 = \frac{3 \kappa}{4} |W_0|^2; \quad \kappa =\left(\frac{g_s}{8\pi}\right).
\eea
Thus using the leading order effects, the derivative of the scalar potential and the Hessian with respect to the overall volume ${\cal V}$ can be given as,
\bea
\label{eq:derV-HessV-leading}
& & \frac{\partial V_1}{\partial {\cal V}} \simeq {\cal C}_1\frac{2 \hat{\eta}\left(7-3\,\ln{\cal V}\right)- 3 \hat{\xi}}{{\cal V}^4} + \cdots,\\
& & \frac{\partial^2 V_1}{\partial {\cal V}^2} \simeq -{\cal C}_1\frac{2 \hat{\eta}\left(31-12\,\ln{\cal V}\right)- 12 \hat{\xi}}{{\cal V}^5} + \cdots,\nonumber
\eea
where $\cdots$ denotes the sub-leading pieces. Subsequently, the extremisation of the scalar potential gives the following constraint,
\bea
\label{eq:pert-LVS-VEV}
& & 2 \hat{\eta}\left(7-3\,\ln{\langle{\cal V}\rangle}\right)\simeq 3 \hat{\xi} \quad \implies \quad \langle {\cal V} \rangle \simeq e^{\frac{14 \hat\eta - 3\hat\xi}{6 \hat\eta}},
\eea
which shows that the overall volume can be dynamically stabilised to exponentially large values by considering small values of string coupling, for a set of given natural values for the $\hat\xi$ and $\hat\eta$ parameters provided that the ratio $\xi/\eta$ is negative. In fact in \cite{Antoniadis:2019doc,Antoniadis:2019rkh,Antoniadis:2020ryh,Antoniadis:2020stf} it has been already found that $\xi/\eta < 0$. For numerical estimates if one takes  $\hat\xi = 8$ and $\hat\eta = -1/2$, one gets $\langle {\cal V} \rangle \simeq 30740$~.

Further, using the extremisation condition (\ref{eq:pert-LVS-VEV}) in (\ref{eq:derV-HessV-leading}) one can eliminate $\hat\xi$ and subsequently can ensure that the solution corresponds to an AdS minimum as seen from the expressions below,
\bea
\label{eq:Vmin-AdS}
& & \hskip-1cm \langle V_1 \rangle \simeq  \frac{2\hat{\eta}\,{\cal C}_1}{3\langle{\cal V}\rangle^3} , \qquad \left\langle \frac{\partial^2 V_1}{\partial {\cal V}^2} \right\rangle \simeq - \frac{6\,\hat{\eta}\,{\cal C}_1}{\langle{\cal V}\rangle^5}.
\eea
Given that ${\cal C}_1 >0$ as seen from Eq.~(\ref{eq:pheno-potV3}) and one needs $\hat\eta < 0$ for a positive VEV of the Hessian component which subsequently corresponds to an AdS minimum, similar to the standard LVS \cite{Balasubramanian:2005zx}. 

In this way, using Eq.~(\ref{eq:def-xi-eta}) and Eq.~(\ref{eq:pert-LVS-VEV}) one finds that the overall volume ${\cal V}$ can be dynamically stabilised to exponentially large values (by using the leading order perturbative effects) in the following manner:
\bea
\label{eq:pert-LVS}
& & \langle {\cal V} \rangle \simeq e^{a/g_s^2 + b}, \qquad a = \frac{\zeta[3]}{2 \zeta[2]} \simeq 0.365381, \quad b = \frac73~\cdot
\eea
As a side remark, let us mention that this AdS minimum obtained in perturbative LVS framework involves two $(\alpha^\prime)^3$ pieces at the leading order: (i). the BBHL piece which is at tree level in $g_s$ series and (ii). the logarithmic loop piece which is at the 1-loop level but still embedded within the same $(\alpha^\prime)^3$ order contributions to the scalar potential. Therefore one may be concerned with the viability of such solutions against higher order terms in the string-loop series. In this regard, let us mention that while this problem can be a priory considered as a possible concern but practically one can easily avoid it by considering small enough region of string coupling, say $g_s \leq {\cal O}(0.3)$ or so. For example, using $g_s = 0.2$ in Eq.~(\ref{eq:pert-LVS}) results in $\langle {\cal V} \rangle = 95593.3$ while $g_s = 0.1$ corresponds to $\langle {\cal V} \rangle = 7.61463 \cdot 10^{16}$. Given that the (un-)known string loop effects of the higher order are anticipated to be further suppressed in powers of CY volume ${\cal V}$ and string coupling $g_s$, so we think that the AdS vacua we have in perturbative LVS scheme should be viable against these corrections in the sense that they will not be washed out, and such sub-leading effects can at most produce a shift into the moduli VEVs or make the minima a bit shallower. In fact these arguments are well demonstrated in table \ref{tab_samplings1} in which one can see that the presence of additional string-loop effects along with the higher derivative $F^4$ corrections can indeed shift the VEVs significantly for the choice of larger $g_s$ values, however the AdS minimum is not washed out. We expect similar things for the other (un-)known sub-leading $\alpha^\prime$ and $g_s$ corrections as well.


\section{A generic formula for the perturbative scalar potential}
\label{sec_gen-formula}

We have used the scalar potential effects of the order ${\cal V}^{-3}$ so far, in order to dynamically fix the overall volume modulus ${\cal V}$. We argue that it is likely to be possible to fix the remaining moduli by using the sub-leading effects, and some well known approaches are already available, for example using sub-leading string-loops effects \cite{Berg:2004ek, Berg:2005ja, Berg:2005yu, Berg:2007wt,Cicoli:2007xp} as compared to the ones we discussed in earlier sections, along with the higher derivative $F^4$ corrections \cite{Ciupke:2015msa}.

\subsection{Sub-leading string-loop effects}

Apart from the leading order string-loop effects which contribute at the same order in the volume scaling as the BBHL correction, there are some other string-loop corrections to the K\"ahler potential which have been computed for toroidal models through various routes \cite{Berg:2004ek, Berg:2005ja, Berg:2005yu, Berg:2007wt}, and have been subsequently conjectured for  generic CY orientifolds \cite{Cicoli:2007xp}. It turns out that the scalar potential is protected against the leading order pieces of such corrections due to the so-called ``extended" no-scale structure \cite{vonGersdorff:2005bf,Cicoli:2007xp} which subsequently results in making these corrections appear at order ${\cal V}^{-10/3}$.  

These additional corrections can be classified into two categories; one is called as the KK-type correction while the other one as winding-type corrections. After a series of works \cite{Berg:2004ek, Berg:2005ja, Berg:2005yu, Berg:2007wt, Cicoli:2007xp}, these corrections have been conjectured to take the following form in the Einstein frame,
\bea
\label{eq:KgsE}
& & \hskip-1cm K_{g_s}^{\rm KK} = g_s \sum_\alpha \frac{C_\alpha^{\rm KK} \, t^\alpha_\perp}{\cal V} \,, \qquad K_{g_s}^{\rm W} =  \sum_\alpha \frac{C_\alpha^W}{{\cal V}\, t^\alpha_\cap}\,,
\eea
where $C_\alpha^{\rm KK}$ and $C_\alpha^{\rm W}$ are some functions which can generically depend on the complex structure moduli and open-string moduli. The two-cycle volume moduli $t^\alpha_{\perp}$ denote the transverse distance among the various stacks of the non-intersecting $D7$-brane and $O7$-planes, whilst  $t^\alpha_{\cap}$ denotes the volume of the curve sitting at the intersection loci of the various non-trivially intersecting stacks of $D7$-branes such that the intersecting curve is non-contractible. This also justifies the appearance of $t^\alpha_{\cap}$ in the denominator as the corresponding curves being non-contractible ensures that it cannot be shrinked to zero size. Some concrete realisations of these Ans\"atze ensuring the string-loop effects have been presented in explicit Calabi-Yau orientifold settings in~\cite{Cicoli:2016xae, Cicoli:2017axo}. The scalar potential contributions arising from these K\"ahler potentials in Eq.~(\ref{eq:KgsE}) are given as~\cite{Cicoli:2007xp}:
\bea
\label{eq:VgsKK-W}
& & V_{g_s}^{\rm KK} = \kappa\, g_s^2 \frac{|W_0|^2}{\vo^2} \sum_{\alpha\beta} C_\alpha^{\rm KK} C_\beta^{\rm KK} K_{\alpha\beta}^0 \,, \\
& & V_{g_s}^{\rm W} = -2 \kappa\, \frac{|W_0|^2}{\vo^2} \, K_{g_s}^{\rm W} = -2 \kappa\, \frac{|W_0|^2}{\vo^3} \, \sum_\alpha \frac{C_\alpha^W}{t^\alpha_\cap}\,; \qquad \kappa = \left(\frac{g_s}{8 \pi}\right), \nonumber
\eea
where $K_{\alpha\beta}^0$ is the tree-level K\"ahler metric which using Eq. (\ref{eq:Pi-tPi-tree}) in Eq. (\ref{eq:Kij}) results in the well known form as given below,
\bea
\label{eq:Kij-tree}
& & K_{\alpha\beta}^0 = \frac{1}{16\, {\cal V}^2}\left(2\,t^\alpha\, t^\beta - 4 {\cal V}\, k^{\alpha\beta} \right).
\eea

\subsection{Higher derivative $F^4$-corrections}
Apart from the BBHL and string-loop corrections, there is a different type  of higher derivative correction which appear at ${\cal O}(F^4)$ in the scalar potential and it is not captured at the level of two-derivative approximations \cite{Ciupke:2015msa}. This correction is argued to be generic for a given Calabi-Yau orientifold compactification and takes the following simple form,
\bea
\label{VF4}
& & V_{F^4} = - \kappa^2 \frac{\lambda\,|W_0|^4}{g_s^{3/2} {\cal V}^4} \Pi_\alpha \, t^\alpha \, 
;\qquad \kappa = \left(\frac{g_s}{8 \pi}\right),
\eea
where $t^\alpha$'s are the volume of the 2-cycles for the generic CY manifold $X$ while $\lambda$ is defined to be a quantity which does not capture any dependence on the volume moduli, and $\Pi_\alpha$ are topological numbers defined as:
\bea
\label{Pii}
& & \Pi_\alpha = \int_X c_2(X) \wedge \hat{D}_\alpha \,.
\eea
Here $c_2(X)$ is the CY second Chern class, $\hat{D}_\alpha$ is a basis of harmonic 2-forms such that the K\"ahler form can be expanded as $J = t^\alpha \,\hat{D}_\alpha$ and $\lambda$ is an unknown combinatorial factor which is expected to be between $10^{-2}$ and $10^{-3}$. 

\subsection{Master formula}
Combining all the perturbative effects collected so far, namely the BBHL's $(\alpha^\prime)^3$ corrections \cite{Becker:2002nn}, the perturbative string-loop effects of \cite{Antoniadis:2018hqy} as well as the higher derivative $F^4$ corrections of \cite{Ciupke:2015msa}, a master formula for perturbative scalar potential using Gukov-Vafa-Witten's flux superpotential $W_0$ can be generically given by the following pieces,
\bea
\label{eq:masterVgen}
& & \hskip-0.9cm V_{\rm pert} = V_{\alpha^\prime + {\rm log} \, g_s} + V_{g_s}^{\rm KK} + V_{g_s}^{\rm W} +V_{F^4} + \dots \\
& & \simeq \frac{\kappa}{{\cal Y}^2} \biggl[\frac{3\, {\cal V} }{2\, {\cal Y}^2} \left(1 + \frac{\partial {\cal Y}_1}{\partial{\cal V}}\right)^2 \left(6 {\cal V} \tilde{\cal P}_3 - \tilde{\cal P}_4 \right) - 3 \biggr]\, |W_0|^2 \nonumber\\
& & +\kappa \, g_s^2 \, \frac{|W_0|^2}{16\,\vo^4} \sum_{\alpha,\beta} C_\alpha^{\rm KK} C_\beta^{\rm KK} \left(2\,t^\alpha t^\beta - 4\, {\cal V} \,k^{\alpha\beta}\right)\nonumber\\
& & -2 \kappa \frac{|W_0|^2}{\vo^3} \, \sum_\alpha \frac{C_\alpha^W}{t^\alpha_\cap} - \kappa^2 \frac{\lambda\,|W_0|^4}{g_s^{3/2} {\cal V}^4} \Pi_\alpha \, t^\alpha + \dots \,, \nonumber
\eea
where just to recall again, we have set $e^{K_{cs}} =1$ and $\kappa = \left(\frac{g_s}{8 \pi}\right)$. Let us note that this master formula of the scalar potential is quite general and can be applied to various possible scenarios. In fact considering the possible contractions in mind we can anticipate that the generic extremisation conditions can be derived to be of the following form,
\bea
\label{eq:derVmain}
& & \partial_{T_\alpha} V = h_1 \, t^\alpha + h_2 \, \Pi_\beta\ \kappa^{\alpha\beta},
\eea
where $h_i$'s are some scalar functions depending on the volume moduli $t^\alpha$ which may involve contractions among quantities like $k^{\alpha\beta}, t^\alpha$ and $k_{\alpha}$, and hence will depend on the overall volume as well. Given the implicit nature of $\tilde{\cal P}_i$'s in terms of the K\"ahler potential ingredients, it is not very illuminating to present the generic form of the $h_1$ and $h_2$ functions. However one should also recall that both the terms in Eq.~(\ref{eq:derVmain}) are not designed to be generically competing which is also desirable as one piece appears at two-derivative ($F^2$) approximation while the other one is generated at $F^4$-order, and therefore, the leading order extremizations may be approximated by the vanishing of the leading order pieces in $h_1$ function alone, at least for some simple form of Ans\"atze  for the string-loop effects entering via ${\cal Y}_1$ or $f({\cal V})$ as we will present later on.


\section{A concrete example for the global model building}
\label{sec_global-model}

In this section we start by presenting an explicit CY threefold example which possesses a toroidal like volume form given as below,
\bea
& & {\cal V} = a \, t^1 \, t^2 \, t^3 = \frac{1}{\sqrt{a}} \sqrt{\tau_1 \, \tau_2 \, \tau_3}
\eea
The main motivation behind the above toroidal looking volume form follows from the proposal of \cite{Antoniadis:2018hqy,Antoniadis:2018ngr,Antoniadis:2019doc,Antoniadis:2019rkh,Antoniadis:2020ryh,Antoniadis:2020stf, Antoniadis:2021lhi} where some symmetries between the various volume moduli were needed for the setting of the overall mechanism. For this purpose, we explored the CY dataset of Kreuzer-Skarke \cite{Kreuzer:2000xy} with $h^{1,1} = 3$ and find that there are at least two geometries which could suitably give this volume form.  One such CY threefold corresponding to the polytope Id: 249 in the CY database of \cite{Altman:2014bfa} can be defined by the following toric data:
\begin{center}
\begin{tabular}{|c|ccccccc|}
\hline
Hyp & $x_1$  & $x_2$  & $x_3$  & $x_4$  & $x_5$ & $x_6$  & $x_7$       \\
\hline
4 & 0  & 0 & 1 & 1 & 0 & 0  & 2   \\
4 & 0  & 1 & 0 & 0 & 1 & 0  & 2   \\
4 & 1  & 0 & 0 & 0 & 0 & 1  & 2   \\   
\hline
& $K3$  & $K3$ & $K3$ &  $K3$ & $K3$ & $K3$  &  SD  \\
\hline
\end{tabular}
\end{center}
The Hodge numbers are $(h^{2,1}, h^{1,1}) = (115, 3)$, the Euler number is $\chi=-224$ and the SR ideal is:
\be
{\rm SR} =  \{x_1 x_6, \, x_2 x_5, \, x_3 x_4 x_7 \} \,. \nn
\ee
This CY threefold was also considered for exploring odd-moduli and exchange of non-trivially identical divisors in \cite{Gao:2013pra}. Moreover, a del-Pezzo upgraded version of this example which corresponds to a CY threefold with $h^{1,1}=4$ has been considered in for chiral global embedding of Fibre inflation model in \cite{Cicoli:2017axo}.

The analysis of the divisor topologies using {\it cohomCalg} \cite{Blumenhagen:2010pv, Blumenhagen:2011xn} shows that they can be represented by the following Hodge diamonds:
\bea
K3 &\equiv& \begin{tabular}{ccccc}
    & & 1 & & \\
   & 0 & & 0 & \\
  1 & & 20 & & 1 \\
   & 0 & & 0 & \\
    & & 1 & & \\
  \end{tabular}, \qquad \quad {\rm SD} \equiv \begin{tabular}{ccccc}
    & & 1 & & \\
   & 0 & & 0 & \\
  27 & & 184 & & 27 \\
   & 0 & & 0 & \\
    & & 1 & & \\
  \end{tabular}.
\eea
Considering the basis of smooth divisors $\{D_1, D_2, D_3\}$ we get the following intersection polynomial which has just one non-zero classical triple intersection number \footnote{There is another CY threefold in the database of \cite{Altman:2014bfa} which has the intersection polynomial of the form $I_3 = D_1\, D_2\, D_3$, however that CY threefold (corresponding to the polytope Id: 52) has non-trivial fundamental group.}:
\bea
& & I_3 = 2\, D_1\, D_2\, D_3,
\eea
while the second Chern-class of the CY is given by,
\bea
c_2(CY) = 5 D_3^2+12 D_1 D_2 + 12 D_2 D_3+12 D_1 D_3.
\eea
Subsequently, considering the K\"ahler form $J = \sum_{\alpha =1}^3 t^\alpha D_\alpha$, the overall volume and the 4-cycle volume moduli can be given as follows:
\bea
& & \hskip-1cm \vo = 2\, t^1\, t^2\, t^3, \qquad \qquad \tau_1 = 2\, t^2 t^3,  \quad  \tau_2 = 2\, t^1 t^3, \quad  \tau_3 = 2 \,t^1 t^2 \,.
\label{Taus}
\eea
This volume form can also be expressed in the following form:
\bea
& & \hskip-1cm {\cal V} = 2 \, t^1\, t^2\, t^3 = t^1 \tau_1 = t^2 \tau_2 = t^3 \tau_3 = \frac{1}{\sqrt{2}}\,\sqrt{\tau_1 \, \tau_2\, \tau_3}~.
\eea
This confirms that the volume form ${\cal V}$ is indeed like a toroidal case with an exchange symmetry $1 \leftrightarrow 2 \leftrightarrow 3$ under which all the three $K3$ divisors which are part of the basis are exchanged. Further, the K\"ahler cone for this setup is described by the conditions below,
\bea
\label{KahCone}
\text{K\"ahler cone:} &&  t^1 > 0\,, \quad t^2 > 0\,, \quad t^3 > 0\,.
\eea
For the classical triple intersection numbers we have, the tree-level K\"ahler metric in Eq. (\ref{eq:Kij-tree}) takes the following form,
\bea
& & K_{\alpha\beta}^0 = \frac{1}{4\, {\cal V}^2} \left(
\begin{array}{ccc}
 (t^1)^2 & 0 & 0 \\
 0 & (t^2)^2 & 0 \\
 0 & 0 & (t^3)^2 \\
\end{array}
\right).
\eea
Now given that there are no rigid divisors present, a priori this setup will not receive non-perturbative superpotential contributions from instanton or gaugino condensation. In fact because of the very same reason this CY could be naively considered to be not well suited for doing phenomenology in the conventional sense, pertaining to the obstacles in stabilising the K\"ahler moduli. However we will show that this is not the case when we include all the perturbative effects arising from the $\alpha^\prime$ as well as $g_s$ series.

Further in order to cancel all D7-charges, we shall introduce $N_a$ D7-branes wrapped around suitable divisors (say $D_a$) and their orientifold images ($D_a^\prime$) such that \cite{Blumenhagen:2008zz}:
\bea
\label{eq:D7tadpole}
& & \sum_a\, N_a \left([D_a] + [D_a^\prime] \right) = 8\, [{\rm O7}]\,.
\eea
D7-branes and O7-planes also give rise to D3-tadpoles which receive contributions also from background 3-form fluxes $H_3$ and $F_3$, D3-branes and O3-planes. The D3-tadpole cancellation condition reads \cite{Blumenhagen:2008zz}:
\be
N_{\rm D3} + \frac{N_{\rm flux}}{2} + N_{\rm gauge} = \frac{N_{\rm O3}}{4} + \frac{\chi({\rm O7})}{12} + \sum_a\, \frac{N_a \left(\chi(D_a) + \chi(D_a^\prime) \right) }{48}\,,
\label{eq:D3tadpole}
\ee
where $N_{\rm flux} = (2\pi)^{-4} \, (\alpha^\prime)^{-2}\int_X H_3 \wedge F_3$ is the contribution from background fluxes and $N_{\rm gauge} = -\sum_a (8 \pi)^{-2} \int_{D_a}\, {\rm tr}\, {\cal F}_a^2$ is due to D7 worldvolume fluxes. For the simple case where D7-tadpoles are cancelled by placing 4 D7-branes (plus their images) on top of an O7-plane, (\ref{eq:D3tadpole}) reduces to:
\be
N_{\rm D3} + \frac{N_{\rm flux}}{2} + N_{\rm gauge} =\frac{N_{\rm O3}}{4} + \frac{\chi({\rm O7})}{4}\,.
\label{eq:D3tadpole1}
\ee
As a consistency check for a given orientifold involution, one has to ensure that the right-hand-side of (\ref{eq:D3tadpole1}) is an integer. 

To begin with looking for the suitable brane settings we note that there are six equivalent reflection involutions corresponding to flipping first six coordinates, i.e. $x_i \to - x_i$ for each $i \in \{1, 2, .., 6\}$. So now let us take the involution $x_1 \to - x_1$, which leads to the fixed point set having two $O7$-plane components given as $\{O7_1 = D_1, \, O7_2 = D_6 \}$ while there are no $O3$-planes present. This results in non-intersecting stacks of $D7$-branes only, and hence cannot produce the string-loop effects of winding-type, although this may induce KK-type string loop effects. However, given the fact that GLSM charges corresponding to $D_1$ and $D_6$ divisors are the same, this situation is like putting a single stack of all the $D7$-branes on top of the $O7$-plane itself, and hence KK-type corrections should not play any r\^ole in the low energy dynamics. Subsequently one can consider the following brane setting involving 2 stacks of $D7$-branes wrapping the divisors $\{D_1, D_6\}$ in the basis,
\bea
& & 8\, [O_7] = 4 \left([D_1] + [D_1^\prime] \right) + 4 \left([D_6] + [D_6^\prime] \right)\,.
\eea
The $D3$ tadpole condition reads as 
\be
N_{\rm D3} + \frac{N_{\rm flux}}{2} + N_{\rm gauge} = 0 + \frac{24+24}{12} + \frac{4(24+24)}{48} + \frac{4(24+24)}{48} = 12\,.
\ee
Although, this involution does help in providing support for non-existence of some of the well known string-loop effects, this does not result in enough scalar potential contributions to have interesting phenomenological implications.

Now considering the involution $x_7 \to - x_7$ leads to the only fixed point set being given as $\{O7 = D_7\}$ as there are no $O3$-planes present. Subsequenlty we consider the following brane setting involving 3 stacks of $D7$-branes wrapping each of the three divisors $\{D_1, D_2, D_3\}$ in the basis,
\bea
& & 8\, [O_7] = 8 \left([D_1] + [D_1^\prime] \right) + 8 \left([D_2] + [D_2^\prime] \right) + 8 \left([D_3] + [D_3^\prime] \right)\,.
\eea
The $D3$ tadpole condition reads as 
\be
N_{\rm D3} + \frac{N_{\rm flux}}{2} + N_{\rm gauge} = 0 + \frac{240}{12} + 8 + 8 + 8 = 44\,,
\ee
which shows some flexibility with turning on background flux as well as the gauge flux. Let us note that the volume form can be given as,
\bea
{\cal V} = t^1 \, \tau_1 = t^2 \, \tau_2  = t^3 \, \tau_3,
\eea
which means that the transverse distance for the stacks of $D7$-branes wrapping the divisor $D_1$ is given by $t^1$ and similarly $t^2$ is the transverse distance for $D7$-branes wrapping the divisor $D_2$ and so on. Moreover the divisor intersection curves are given in table \ref{Tab1} which shows that all the three $D7$-brane stacks intersect at ${\mathbb T}^2$ while each of those intersect the $O7$-plane on a curve ${\cal H}_9$ defined by $h^{0,0} = 1$ and $h^{1,0} = 9$. These properties about the transverse distances and the divisor interesting on ${\mathbb T}^2$ is perfectly like what one has for the toroidal case, though the divisors are $K3$ for the current situation unlike ${\mathbb T}^4$ divisors of the six-torus. These symmetries are consistent with the basis requirement for generating logarithmic string-loop effects as elaborated in \cite{Antoniadis:2018hqy,Antoniadis:2018ngr,Antoniadis:2019doc,Antoniadis:2019rkh,Antoniadis:2020ryh,Antoniadis:2020stf,Antoniadis:2021lhi}. 

Further we note that there are no non-intersection $D7$-brane stacks and the $O7$-planes along with no $O3$-planes present as well, and therefore this model does not induce the KK-type string-loop corrections to the K\"ahler potential. However, given the fact that $D7$-brane stacks intersection on non-shrinkable two-torus, one will have string-loop effects of the winding-type to be given as below,
\bea
\label{eq:Vgs-Winding-globalmodel}
& & V_{g_s}^{\rm W} = - \left(\frac{g_s}{8 \pi}\right) \frac{|W_0|^2}{\vo^3} \, \sum_{\alpha=1}^3 \frac{C_\alpha^W}{t^\alpha}\,,
\eea
Note that we have used the volume of a given two-torus $t^\alpha_\cap$ at the intersection locus of any two $D7$-brane stacks as given below,
\bea
& & \hskip-1.5cm \int_{CY} J \wedge D_1 \wedge D_2 = 2t^3, \quad \int_{CY} J \wedge D_2 \wedge D_3 = 2t^1, \quad \int_{CY} J \wedge D_3 \wedge D_1 = 2t^2,
\eea
where the K\"ahler form is taken as $J = t^1 D_1 + t^2 D_2 + t^3 D_3$.

\begin{table}[h]
  \centering
 \begin{tabular}{|c|c|c|c|c|c|c|c|}
\hline
  & $D_1$  & $D_2$  & $D_3$  & $D_4$  & $D_5$ & $D_6$  & $D_7$  \\
    \hline
		\hline
$D_1$ & $\emptyset$  &  ${\mathbb T}^2$      &  ${\mathbb T}^2$        &  ${\mathbb T}^2$   &  ${\mathbb T}^2$  &  $\emptyset$   &  ${\cal H}_9$  \\
$D_2$ &  ${\mathbb T}^2$ & $\emptyset$        &  ${\mathbb T}^2$        &  ${\mathbb T}^2$   &    $\emptyset$   & ${\mathbb T}^2$  & ${\cal H}_9$
\\
$D_3$  &  ${\mathbb T}^2$      &  ${\mathbb T}^2$        & $\emptyset$ & $\emptyset$ &  ${\mathbb T}^2$   &  ${\mathbb T}^2$   &  ${\cal H}_9$
\\
$D_4$  &  ${\mathbb T}^2$      &  ${\mathbb T}^2$        & $\emptyset$ & $\emptyset$ &  ${\mathbb T}^2$   &  ${\mathbb T}^2$   &  ${\cal H}_9$
\\
$D_5$ &  ${\mathbb T}^2$ & $\emptyset$        &  ${\mathbb T}^2$        &  ${\mathbb T}^2$   &    $\emptyset$   & ${\mathbb T}^2$  & ${\cal H}_9$
\\
$D_6$ & $\emptyset$  &  ${\mathbb T}^2$      &  ${\mathbb T}^2$        &  ${\mathbb T}^2$   &  ${\mathbb T}^2$  &  $\emptyset$   &  ${\cal H}_9$  \\
$D_7$ & ${\cal H}_9$  &  ${\cal H}_9$      &  ${\cal H}_9$        &  ${\cal H}_9$   &  ${\cal H}_9$  &  ${\cal H}_9$   &  ${\cal H}_{97}$
\\
    \hline
  \end{tabular}
  \caption{Intersection curves of the two coordinate divisors. Here ${\cal H}_g$ denotes a curve with Hodge numbers $h^{0,0} = 1$ and $h^{1,0} = g$, and hence ${\cal H}_1 \equiv {\mathbb T}^2$, while ${\cal H}_0 \equiv {\mathbb P}^1$.} 
\label{Tab1}
\end{table}

\begin{table}[h]
  \centering
 \begin{tabular}{|c|c|c|c|c|c|c|c|}
\hline
  & $D_1$  & $D_2$  & $D_3$  & $D_4$  & $D_5$ & $D_6$  & $D_7$  \\
    \hline
		\hline
$D_1$  & 0 & 2 ${t^3}$ & 2 ${t^2}$ & 2 ${t^2}$ & 2 ${t^3}$ & 0 & 4 ${t^2}$+4 ${t^3}$ \\
$D_2$  & 2 ${t^3}$ & 0 & 2 ${t^1}$ & 2 ${t^1}$ & 0 & 2 ${t^3}$ & 4 ${t^1}$+4 ${t^3}$ \\
$D_3$  & 2 ${t^2}$ & 2 ${t^1}$ & 0 & 0 & 2 ${t^1}$ & 2 ${t^2}$ & 4 ${t^1}$+4 ${t^2}$ \\
$D_4$  & 2 ${t^2}$ & 2 ${t^1}$ & 0 & 0 & 2 ${t^1}$ & 2 ${t^2}$ & 4 ${t^1}$+4 ${t^2}$ \\
$D_5$  & 2 ${t^3}$ & 0 & 2 ${t^1}$ & 2 ${t^1}$ & 0 & 2 ${t^3}$ & 4 ${t^1}$+4 ${t^3}$ \\
$D_6$  & 0 & 2 ${t^3}$ & 2 ${t^2}$ & 2 ${t^2}$ & 2 ${t^3}$ & 0 & 4 ${t^2}$+4 ${t^3}$ \\
$D_7$  & 4 ${t^2}$+4 ${t^3}$ & 4 ${t^1}$+4 ${t^3}$ & 4 ${t^1}$+4 ${t^2}$ & 4 ${t^1}$+4 ${t^2}$ & 4 ${t^1}$+4
   ${t^3}$ & 4 ${t^2}$+4 ${t^3}$ & 16 $(t^1 + t^2 + t^3)$ \\
    \hline
  \end{tabular}
  \caption{Volume of the two-cycles at the intersection local of the two coordinate divisors $D_i$ presented in Table \ref{Tab1}. This shows, for example, that the curve intersecting at divisors $D_1$ and $D_2$ has a volume along $t^3$, like in the usual toroidal scenarios.} 
\label{Tab2}
\end{table}

\noindent
Finally let us note that the topological quantities $\Pi_i$'s appearing in the higher derivative $F^4$ corrections are given as,
\bea
& & \Pi_i = 24 \quad \forall \, i \in \{1, 2,..,6\}; \quad \Pi_7 = 124.
\eea
Thus,  we observe that although this CY have several properties like a toroidal case, the divisor being $K3$ implies their corresponding $\Pi = 24$ unlike the ${\mathbb T}^4$ case which has a vanishing $\Pi$, and hence no such higher derivative effects. Subsequently we find the following scalar potential corrections,
\bea
\label{eq:F^4-term-globalmodel}
& & V_{F^4} = - \left(\frac{g_s}{8 \pi}\right)^2 \frac{\lambda\,|W_0|^4}{g_s^{3/2} {\cal V}^4}\, 24 \, \left(t^1 + t^2 + t^3\right)
\eea


\section{Moduli stabilisation and de Sitter vacua}
\label{sec_moduli-stabilization}

We take a two step strategy to do moduli stabilisation, first in an AdS vacuum and then uplift the same to de Sitter via means to adding $D$-term effects. 

\subsection{Fixing all moduli in perturbative LVS}
With the master formula and simplified versions derived so far, we are now in a position to perform the study of moduli stabilisation. For this purpose, the main idea is to fix the overall volume by the leading order ${\cal O}({\cal V}^{-3})$ terms arising from BBHL and logarithmic-loop effects, and then fix the remaining moduli by a combination of winding string-loop effects at order ${\cal O}({\cal V}^{-10/3})$ and some more volume suppressed terms, e.g. the higher order $F^4$ effects with volume scaling  ${\cal O}({\cal V}^{-11/3})$ etc. So we consider the simplified version of the scalar potential in Eq. (\ref{eq:masterVgen}) given as below,
\bea
\label{eq:Vfinal-simp}
& & \hskip-1cm V \simeq {\cal C}_1 \left(\frac{\hat\xi - 4\,\hat\eta + 2\,\hat\eta \, \ln{\cal V}}{{\cal V}^3}\right) + 6 \, {\cal C}_1 \left(\frac{3 \hat{\eta } \hat{\xi }+4 \hat{\eta }^2+\hat{\xi }^2-2 \hat{\eta } \hat{\xi} \ln{\cal V} -2 \hat{\eta }^2 \, \ln{\cal V}}{{\cal V}^4}\right) \\
& & + \frac{{\cal C}_2}{{\cal V}^3} \left(\frac{1}{t^1} + \frac{1}{t^2}+\frac{1}{t^3} \right) +  \frac{{\cal C}_3\,\left(t^1 + t^2 + t^3\right)}{{\cal V}^4} + {\cal O}({\cal V}^{-5}) + \dots \,, \nonumber
\eea
where the first line descends from our master formula in Eq.~(\ref{eq:masterV}), and first piece in the second line corresponds to the winding-type string loop effects as collected in  Eq.~(\ref{eq:Vgs-Winding-globalmodel}) while the second piece in the second line corresponds to the higher derivative $F^4$ effects as given in Eq.~(\ref{eq:F^4-term-globalmodel}). Subsequently the coefficients ${\cal C}_i$'s are given as below:
\bea
\label{eq:calCis}
& & \hskip-1cm {\cal C}_1 = \left(\frac{g_s}{8\pi}\right)\frac{3\,|W_0|^2}{4}, \qquad {\cal C}_2 =  \left(\frac{g_s}{8 \pi}\right)|W_0|^2, \qquad {\cal C}_3 = - \left(\frac{g_s}{8 \pi}\right)^2 \frac{\lambda\,|W_0|^4}{g_s^{3/2}}~\cdot
\eea
Note that we have set $e^{K_{cs}} =1$, and we set $C_1^W = C_2^W = C_3^W = -1$. For the current global model candidate, the Euler characteristic is: $\chi(CY) = -224$, and using Eq.~(\ref{eq:def-xi-eta}) we have,
\bea
& & \hskip-1cm \hat\xi 
= \frac{14\, \zeta[3]}{\pi^3\, g_s^{3/2}}, \qquad \hat\eta 
= - \frac{14 \sqrt{g_s} \zeta[2]}{\pi^3}, \qquad \frac{\hat\xi}{\hat\eta} = -\frac{\zeta[3]}{\zeta[2]\, g_s^2}\,.
\eea
Note that the approximate VEV of the overall volume ${\cal V}$ can still be given as in Eq.(\ref{eq:pert-LVS}) which has been obtained by using the leading order terms in the large volume approximation and minimising the potential in  the weak coupling regime. 
In fact, using the scalar potential in Eq~(\ref{eq:Vfinal-simp})-(\ref{eq:calCis}), it is possible to stabilise the overall volume by the first term which is leading order with ${\cal O}({\cal V}^{-3})$ while the remaining K\"ahler moduli can be stabilised by the combination of winding-type string-loop effects and the higher derivative $F^4$ corrections appearing at ${\cal O}({\cal V}^{-10/3})$ and ${\cal O}({\cal V}^{-11/3})$ respectively. We have performed a numerical analysis for a collective stabilisation of all the three moduli $t^\alpha$ by using the full scalar potential in Eq~(\ref{eq:Vfinal-simp}) and the relevant details are given in table \ref{tab_samplings1}.

\begin{table}[H]
\begin{center}
\begin{tabular}{|c||c|c|c|c||c|c|c||} 
\hline
&&&&&&&\\
Sample & $g_s$  & $\hat\xi$ & $\hat\eta$ & $\langle{\cal V}_{\rm approx}\rangle$ & $\langle t^\alpha \rangle$ & $\langle {\cal V} \rangle$ & -$\langle V \rangle$ \\
&&&&&&& \\
\hline
&&&&&&&\\
{\bf S1} & 0.10 & 17.1634 & -0.234870 & 7.615$\cdot10^{16}$ & 336422.0 & 7.61522$\cdot10^{16}$ & 1.06$\cdot 10^{-52}$ \\
{\bf S2} & 0.12 & 13.0566 & -0.257287 & 1.07899$\cdot10^{12}$ & 8143.59 & 1.08014$\cdot 10^8$ & 4.88$\cdot 10^{-40}$ \\
{\bf S3} & 0.14 & 10.3612 & -0.277902 & 1.28657$\cdot 10^9$ & 865.908 & 1.29851$\cdot 10^9$ & 3.55$\cdot 10^{-31}$ \\
{\bf S4} & 0.16 & 8.48054 & -0.297089 & 1.62898$\cdot 10^7$  & 203.686 & 1.6901$\cdot 10^7$ & 1.98$\cdot 10^{-25}$ \\
{\bf S5} & 0.18 & 7.10714 & -0.315111 & 814671.0  & 76.479 & 894656.0 & 1.62$\cdot 10^{-21}$ \\
{\bf S6} & 0.20 & 6.06818 & -0.332156 & 95594.5  & 38.553 & 114605.0 & 9.29$\cdot 10^{-19}$ \\
{\bf S7} & 0.25 & 4.34204 & -0.371362 & 3566.85  & 14.2864 & 5831.68 & 1.10$\cdot 10^{-14}$ \\
{\bf S8} & 0.30 & 3.30310 & -0.406806 & 597.723  & 8.73428 & 1332.63 & 1.34$\cdot 10^{-12}$ \\
{\bf S9} & 0.35 & 2.62121 & -0.439401 & 203.576  & 6.56959 & 567.081 & 2.36$\cdot 10^{-11}$ \\
&&&&&&&\\
\hline
\end{tabular}
\end{center}
\caption{Benchmark models presented for a range of values of string coupling. Other parameters are taken as: $W_0 =1, \, C_1^W = C_2^W = C_3^W = -1$ and $\lambda = -0.01$.}
\label{tab_samplings1}
\end{table}

\noindent
From table \ref{tab_samplings1} we note that the approximated values of the CY volume ${\cal V}$ at the minimum receives significant corrections from the sub-leading sources arising from the winding-type and $F^4$-type effects. However the corresponding shifts in ${\cal V}$ are much smaller for smaller values of the string coupling, the reason being the fact that smaller string coupling corresponds to larger volume with an exponential growth in the inverse square of $g_s$, as can be seen from Eq.~(\ref{eq:pert-LVS}). Moreover we also note that the second piece of Eq.~(\ref{eq:Vfinal-simp}) (which is sub-leading as compared to the first term) does not depend on any other K\"ahler moduli except the overall volume, and hence the remaining moduli can be stabilised by the string-loops and higher derivative $F^4$-terms similar to \cite{Ciupke:2015msa,Cicoli:2016chb}. In fact, using the generic CY orientifolds, one should be able to possibly fix all the remaining $(h^{1,1}_+ - 1)$ K\"ahler moduli through the sub-leading effects from BBHL, string-loops and $F^4$-corrections.

\subsection{On de Sitter uplifting}
Let us note that in our present concrete CY construction the choice of orientifold involution which leads to having three stacks of $D7$-branes intersecting at three ${\mathbb T}^2$'s is such that there are no $O3$-planes present, and therefore anti-$D3$ uplifting proposal of \cite{Crino:2020qwk} is not directly applicable to our case. However, there can be various other ways of inducing uplifting term which can result in de-Sitter solution. In this regard we consider the $D$-term potential associated with the anomalous $U(1)$'s living on the stack of $D7$-branes wrapping the $O7$-planes (say corresponding to divisor class $D_h$), which can be expressed as below,
\bea
\label{eq:}
& & V_{D} = \frac{1}{2\, Re(f_{D7})} \left(\sum_i q_{\varphi_i} \frac{|\varphi_i|^2}{Re(S)} - \xi_{h}\right)^2.
\eea
Here $f_{D7} = T_h/(2\pi)$ denotes the holomorphic gauge kinetic function expressed in terms of complexfied four-cycle volume of the divisor $D_h$, and $q_{\varphi_i}$ denotes the $U(1)$ charge corresponding to the matter field $\varphi_i$. These may correspond to, for example, the deformation of divisors wrapped by the respective $D7$-brane stacks and hence can be counted by $h^{2,0}(D)$ of a suitable divisor of the CY threefold. The FI-parameters $\xi_{h}$ are defined as,
\bea
\label{eq:FI-parameter}
& & \xi_{h} = \frac{1}{4\pi\,{\cal V}} \int_{D_h} {\cal F} \wedge J = \frac{1}{2\pi} \sum_\alpha \frac{q_{h\alpha}}{2} \frac{t^\alpha}{{\cal V}} = - \frac{i}{2\pi} \sum_\alpha q_{h\alpha} \frac{\partial K}{\partial T_\alpha},
\eea
where in the last equality the K\"ahler derivatives have been introduced according to the Eq.~(\ref{eq:derK}). Moreover, ${\cal F}$ denotes the gauge flux turned on the Divisor class $D_h$, and $J$ is the K\"ahler form expressed as $J = t^1 D_1 + t^2 D_2 + t^3 D_3$. The $U(1)$ charge corresponding to the closed string modulus $T_\alpha$ is denoted as $q_{h\alpha}$ and can be given as below,
\bea
& & q_{h\alpha} = \frac{1}{l_s^4}\int_{D_h} \hat{D}_\alpha \wedge {\cal F},
\eea
where $l_s$ denoted the string length parametrised as: $l_s = 2\pi \sqrt{\alpha^\prime}$.

For realising de-Sitter solution we present two scenarios; one in which we introduce $D$-term uplifting via Fayet-Iliopoulos term assuming that matter fields receive vanishing VEVs and the second one being a scenario of the so-called $T$-brane uplifting  in which matter field have non-zero VEVs \cite{Cicoli:2015ylx}. Both of these scenarios present a different volume scaling in the scalar potential term inducing the uplifting of the AdS we have realised before.

\subsubsection{Scenario 1: $D$-term uplifting via matter fields of vanishing VEVs}
\begin{figure}[!htp]
\begin{center}
\begin{tikzpicture}[scale=0.9]
\node (plt) at (0,0) {\includegraphics[width=\textwidth,keepaspectratio]{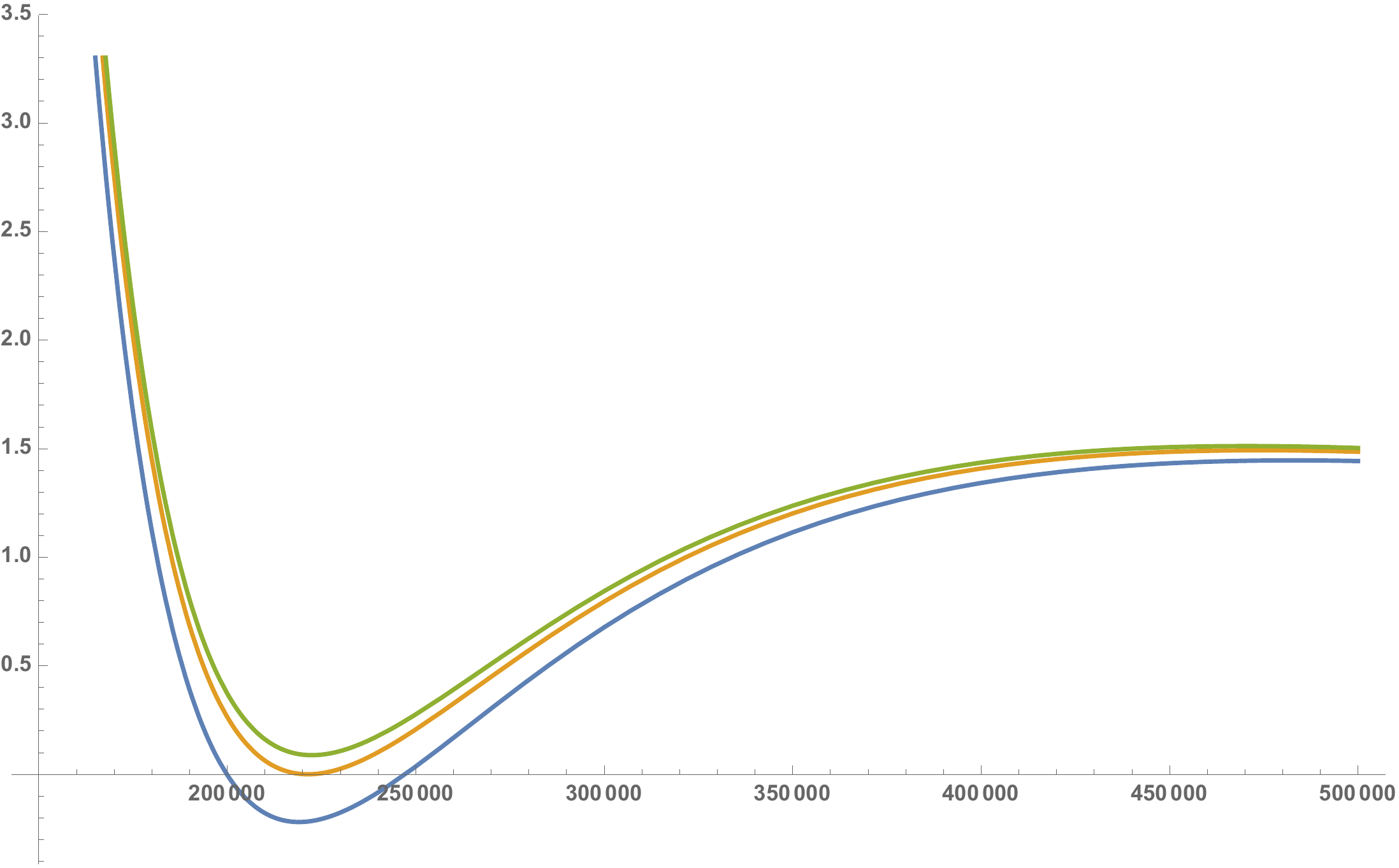}};
\node[above= of plt,node distance=0cm,yshift=-0.75cm,xshift=-7.2cm] {$V({\cal V})$};
\node[below= of plt,node distance=0cm,yshift=1.5cm,xshift=0.75cm] {${\cal V}$};
\node[right= of plt,node distance=0cm,yshift=3.23cm,xshift=-5cm] {dS : $d^\prime = 1.25$};
\node[right= of plt,node distance=0cm,yshift=2.59cm,xshift=-5cm] {Mink. : $d^\prime = 1.24711$};
\node[right= of plt,node distance=0cm,yshift=1.96cm,xshift=-5cm] {AdS: $d^\prime = 1.24$};
\end{tikzpicture}
\end{center}
\caption{The uplifted scalar potential $V({\cal V})$ plotted for overall volume ${\cal V}$. Here $d^\prime = 10^8\, d$ and scalar potential is scaled with a factor $10^{20}$.}
\label{fig1}
\end{figure}

Assuming that matter field receive vanishing VEVs along with each one of the $D7$-brane stack being appropriately magnetised by suitable gauge fluxes so to that generate a moduli-dependent Fayet-Iliopoulos term, one can generically have the following $D$-term contributions to the scalar potential,
\bea
& & V_D \propto \sum_{\alpha =1}^3 \left[\frac{1}{\tau_\alpha} \left(\sum_{\beta \neq \alpha}\, q_{\alpha\beta}\, \frac{\partial K}{\partial \tau_\beta}\right)^2 \right] \simeq \sum_{\alpha =1}^3 \frac{d_\alpha}{f_\alpha^{(3)}},
\eea
where $f_\alpha^{(3)}$ denotes some homogeneous cubic polynomial in generic four-cycle volume $\tau_\beta$. However, given the underlying toroidal-like symmetry the chosen CY threefold possesses it is easy to naturally enforce this symmetry in the moduli VEVs as well, via choosing parameters in a symmetric manner. For example, given the $t^1 \leftrightarrow t^2 \leftrightarrow t^3$ symmetry of the $F$-term scalar potential (\ref{eq:Vfinal-simp}), we can take the model dependent parameters $d_\alpha$ as $d_1 = d_2 = d_3 \equiv d$ which self consistently leads to $\tau_1 = \tau_2 = \tau_3 \equiv \tau$, i.e. $f_\alpha^{(3)} \sim \tau^3$, and hence can facilitate an uplifting of the AdS vacua as presented in table \ref{tab_samplings1} to some de Sitter vacua a la~\cite{Antoniadis:2018hqy}. Subsequently we find that adding this $D$-term effects to the previously analysed scalar potential given in Eqs. (\ref{eq:Vfinal-simp})-(\ref{eq:calCis}), one can indeed have de Sitter uplifting. In figure \ref{fig1} we present the AdS solutions along with its uplifted Minkowskian and de Sitter version for sampling {\bf S6} of table \ref{tab_samplings1}. In this case the numerical parameters and the moduli VEVs corresponding to nearly Minkowskian minimum can be given as below:
\bea
& & g_s = 0.2, \quad \hat\xi = 6.06818, \quad \hat\eta = -0.332156, \quad d = 1.24711\cdot 10^{-8}, \, \nonumber\\
& & \langle t^\alpha \rangle = 48.0191 \, \quad \forall \alpha, \quad \langle {\cal V} \rangle = 221447.96, \quad \langle V \rangle = 1.54074\cdot10^{-31}, \nonumber\\
& & {\rm Eigen}(V_{ij}) = \{5.04286\cdot 10^{-22},\, 5.04286\cdot 10^{-22}, \, 5.04286\cdot 10^{-22}\}.
\eea

\noindent

\subsubsection{Scenario 2: $T$-brane uplifting via matter fields of non-vanishing VEVs}
\begin{figure}[!htp]
\begin{center}
\begin{tikzpicture}[scale=1]
\node (plt) at (0,0) {\includegraphics[width=\textwidth,keepaspectratio]{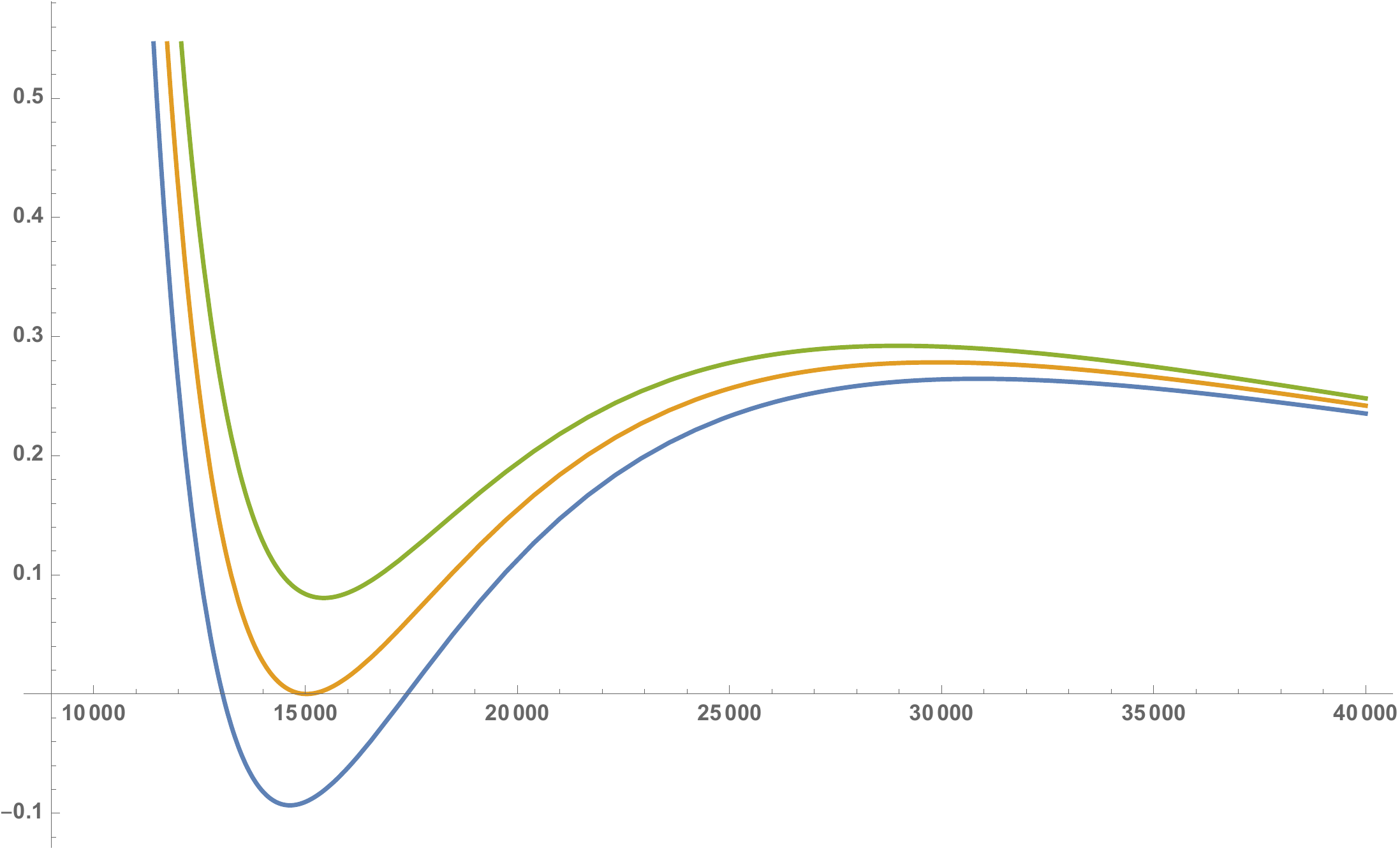}};
\node[above= of plt,node distance=0cm,yshift=-0.75cm,xshift=-7.2cm] {$V({\cal V})$};
\node[below= of plt,node distance=0cm,yshift=1.5cm,xshift=0.75cm] {${\cal V}$};
\node[right= of plt,node distance=0cm,yshift=3.23cm,xshift=-5cm] {dS : ${\cal C}_{up} = 0.0815$};
\node[right= of plt,node distance=0cm,yshift=2.59cm,xshift=-5cm] {Mink. : ${\cal C}_{up} = 0.0814039...$};
\node[right= of plt,node distance=0cm,yshift=1.96cm,xshift=-5cm] {AdS: ${\cal C}_{up} = 0.0813$};
\end{tikzpicture}
\end{center}
\caption{The uplifted scalar potential $V({\cal V})$ plotted for overall volume ${\cal V}$. Here the scalar potential is scaled with a factor $10^{16}$.}
\label{fig2}
\end{figure}
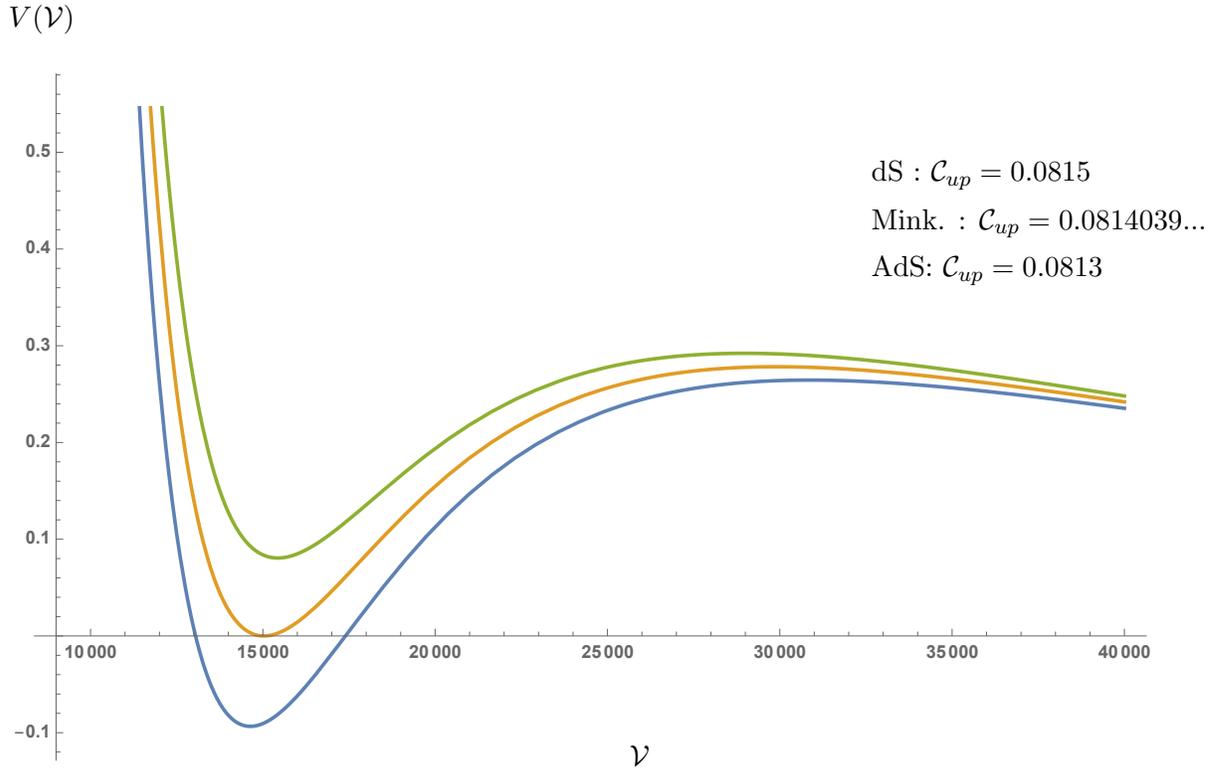

In the presence of non-zero gauge flux on the hidden sector D7-branes, a non-vanishing Fayet-Iliopoulos term can be induced leading to the so-called $T$-brane configuration. It has been shown in \cite{Cicoli:2015ylx} that after expanding the D7-brane action around such $T$-brane background with three-form supersymmetry breaking fluxes, one can get a positive definite uplifting piece to the scalar potential.
Considering such an $T$-brane uplifting case, matter field $\varphi$ receive VEVs of the following kind \cite{Cicoli:2015ylx,Cicoli:2017shd,Cicoli:2021dhg},
\bea
& & |\varphi|^2 \simeq \frac{c_\varphi}{{\cal V}^{2/3}},
\eea
where $c_\varphi$ is a model depending quantity which involves $U(1)$ charges corresponding to the matter fields. This subsequently leads to an uplifting term to the scalar potential induced as a hidden sector supersymmetry breaking $F$-term contribution achieved through the $D$-term stabilization of the matter field, and subsequently the soft-term arising as $F$-term effect can be given as \cite{Cicoli:2015ylx,Cicoli:2017shd,Cicoli:2021dhg},
\bea
\label{eq:T-brane-pot}
& & \hskip-1cm V_{\rm up} = m_{3/2}^2 |\varphi|^2= \left(\frac{g_s}{8 \pi}\right)\,\frac{{\cal C}_{up}\, |W_0|^2}{{\cal V}^{8/3}} \geq 0,
\eea
where $m_{3/2}$ denotes the gravitino mass, and ${\cal C}_{up}$ denotes a model dependent coefficient which also involves the $U(1)$ charges corresponding to the matter fields. One can subsequently use this positive semidefinite piece to uplift the AdS solution of the perturbative LVS to a de-Sitter minimum. In this case, using the uplifting piece in Eq.~(\ref{eq:T-brane-pot}) the numerical parameters and the moduli VEVs corresponding to nearly Minkowskian minimum corresponding to the sampling {\bf S8} of table \ref{tab_samplings1} can be given as below:
\bea
& & g_s = 0.3, \quad \hat\xi = 3.3031, \quad \hat\eta = -0.406806, \quad {\cal C}_{up} = 0.0814039, \, \nonumber\\
& & \langle t^\alpha \rangle = 19.5862 \, \quad \forall \alpha, \quad \langle {\cal V} \rangle = 15027.3, \quad \langle V \rangle = 3.74709\cdot10^{-30} \nonumber\\
& & {\rm Eigen}(V_{ij}) = \{6.81793\cdot 10^{-18},\, 4.68145\cdot 10^{-19}, \, 4.68145\cdot 10^{-19}\}.
\eea
In figure \ref{fig2} we present the modified version of the AdS solution along with its $T$-brane uplifted Minkowskian and de Sitter version for a set of ${\cal C}_{up}$ values. We note that the typical values for this parameter can be  around ${\cal C}_{up} \simeq {\cal O}(0.1)$ as reported in \cite{Cicoli:2017shd,Cicoli:2021dhg}, and this is the reason why we report the sampling {\bf S8} corresponding to $g_s = 0.3$. For lower values of $g_s$ the required value of ${\cal C}_{up}$ parameter turns out to be quite small, say smaller than $0.01$. Moreover we also observe that the including the $T$-brane uplifting term as given in Eq.~(\ref{eq:T-brane-pot}) helps in realizing a larger volume for a relatively larger string coupling regime as compared to what can be typically realized in perturbative LVS condition (\ref{eq:pert-LVS-VEV}). 

\noindent

\noindent
Finally let us mention that we have analysed the uplifting of the full scalar potential resulting in the perturbative LVS using Eq.~(\ref{eq:Vfinal-simp}), which not only includes the leading order BBHL and log-loop pieces but also the subleading terms arising from the sources such as Winding-type string loop effects as well as higher derivative $F^4$ corrections, and therefore one does not need to perform a modified uplifting given that it is not only the overall volume modulus ${\cal V}$ but all the three K\"ahler moduli which are included in the overall numerical dynamics.


\section{Conclusions}
\label{sec_conclusions}

In this work, the  moduli stabilisation  problem has been re-examined  in the framework of type IIB string theory.
 As is well-known, tackling this issue requires the inclusion of quantum contributions in the K\"ahler potential  beyond the classical level. 	The main objective of this article is to investigate this issue  taking into account only perturbative contributions and  consider their  implications in the effective field theory limit. In this context, we have parametrised the string loop effects in terms of a generic function and derived a master formula for  the scalar potential including  also  the contribution of  $\alpha'$ corrections.  We exemplified this generic formula by several paradigms including the case of logarithmic corrections which appear due to local tadpoles induced by the localised gravity kinetic terms stemming from the reduction of $R^4$ terms of  the effective ten-dimensional string action. Subsequently, we further considered a generalisation of our master formula for the scalar potential by including sub-leading string loop corrections which -according to recent conjectures- appear to be generic in Calabi-Yau compactifications. Finally, the implications of a 
 higher derivative effects appearing at order ${\cal O}(F^4)$ in the scalar potential have been considered. Therefore, our most general expression for the scalar 
 potential includes all the essential perturbative  corrections in type IIB string theory and consequently can be applied to a wide class of models. 
 
 We illustrated the presence of various such (sub-)leading corrections to the scalar potential in a concrete $K3$-fibred CY orientifold example chosen from the Kreuzer-Skarke dataset with $h^{1,1}=3$.  This CY threefold is quite unique in the sense that it possesses several properties like those of a toroidal orientifold, say for example ${\mathbb T}^6/({\mathbb Z}_2 \times {\mathbb Z}_2)$ case, such that the volume ${\cal V}$ is given by the product of three 2-cycle moduli $t^\alpha$, i.e., ${\cal V}\propto t^1\, t^2\, t^3$. Moreover intersection of the three $K3$-divisors which are wrapping the three stacks of $D7$-branes intersect on three ${\mathbb T}^2$'s similar to the toroidal case. In this explicit construction we have found that only winding-type string loop corrections can be generated while the KK-type corrections are absent because there are neither non-intersecting $D7/O7$ stacks nor any $O3$-planes which are needed for inducing such corrections. However, unlike the toroidal case, this CY has $K3$ surfaces as divisors unlike the ${\mathbb T}^4$ divisors of the six-torus, and therefore the topological quantity $\Pi(D) = \int_{CY} c_2(CY) \wedge \hat{D}$ characterising the $F^4$ corrections are non-trivially present as well. Using all these corrections, first we have shown that one can have a perturbative large volume minimum realised at exponentially large values in the weak coupling regime. Subsequently the remaining two K\"ahler moduli are lifted by the combined effects of sub-leading corrections arising from the BBHL, string-loops and the Higher derivative $F^4$-corrections. This way we have stabilised all the K\"ahler moduli in a perturbative LVS framework which turns out to be an AdS minimum. Finally, we have discussed possible Fayet-Iliopoulos D-terms associated with the intersecting D7 brane stacks magnetised with gauge fluxes which could generate the necessary uplifting terms to obtain de Sitter vacua in two possible scenarios, first using vanishing VEVs for matter fields and the second being the so-called $T$-brane uplifting method.


\acknowledgments

The work of GKL is supported by the Hellenic Foundation for Research and Innovation (H.F.R.I.) under the ``First Call for H.F.R.I. Research Projects to support Faculty members and Researchers and the procurement of high-cost research equipment
grant'' (Project Number: 2251). PS would like to thank Paolo Creminelli, Atish Dabholkar and Fernando Quevedo for their support.




\bibliographystyle{JHEP}
\bibliography{reference}

\end{document}